\documentclass[twocolumn,showpacs,nofootinbib,preprintnumbers,amsmath,amssymb,showkeys,superscriptaddress]{revtex4-1}
\usepackage{epsfig}
\usepackage{dsfont}
\usepackage{color}
\usepackage{graphicx}
\usepackage{dcolumn}
\usepackage{bm}
\usepackage{physics}

\begin{document}
\allowdisplaybreaks
\newcommand\thetau{\underline{\theta}}

\title{Unequal time correlators of stochastic scalar fields in de Sitter space}

\author{G. Moreau} 
\author{J. Serreau}
\affiliation{APC, AstroParticule et Cosmologie, Universit\'e Paris Diderot, CNRS/IN2P3, CEA/Irfu, Observatoire de Paris, Sorbonne Paris Cit\'e, 10, rue Alice Domon et L\'eonie Duquet, 75205 Paris Cedex 13, France.}
\date{\today}

\begin{abstract}
    The quantum fluctuations of a test scalar field on superhorizon scale in de Sitter spacetime can be described by an effective one-dimensional stochastic theory corresponding to a particular class of nonequilibrium dynamical systems known as the model A. Using the formulation of the latter in terms of a supersymmetric field theory, we compute various unequal time correlators at large (superhorizon) time separations and compare with existing quantum field theory computation. This includes perturbative calculations, pushed here up to three-loop order, and a nonperturbative $1/N$ expansion at next-to-leading order. Exploiting the supersymmetry of the stochastic theory, we also derive a spectral representation of the field correlators and a fluctuation-dissipation relation for the infrared modes of the scalar field in de Sitter spacetime. 
\end{abstract}

\maketitle

\section{Introduction}
\label{sec:intro}

The dynamics of quantum fields in an expanding spacetime is a subject of primordial importance for cosmology and inflation. In this context, the usual approach is a semi-classical treatment where spacetime is treated classically and interacts with a matter content which is of quantum nature and can have a possible backreaction on the geometry \cite{Birrell:1982ix}. De Sitter spacetime is of particular interest both physically, as it is a good approximation of the inflationary phase, and mathematically, because of its high degree of symmetry. 

The computation of quantum corrections in the presence of interactions is a lot more complicated in a curved background as usual perturbative tools are not always available. One particular setup in which nontrivial effects arise is the case of light scalar fields in the expanding Poincar\'e patch of de Sitter spacetime, particularly relevant for inflationary cosmology. The scalar fields mode functions are significantly modified by the curvature with, in particular, a strong amplification of the infrared modes, which can be viewed as intense particle production from the gravitational field \cite{Mottola:1984ar,Tsamis:2005hd,Krotov:2010ma}. This effect is at the origin of infrared and secular divergences in loop computations that limit the use of perturbation theory \cite{Weinberg:2005qc,Starobinsky:1994bd}. 

A variety of nonperturbative treatments exists to address the question of the nonlinear effects ({\it e.g.} self-interactions), see Refs.~\cite{Starobinsky:1994bd,Tsamis:2005hd,vanderMeulen:2007ah,Burgess:2009bs,Rajaraman:2010xd,Beneke:2012kn,Serreau:2011fu,Akhmedov:2011pj,Garbrecht:2011gu,Boyanovsky:2012qs,Parentani:2012tx,Kaya:2013bga,Serreau:2013eoa,Gautier:2013aoa,Youssef:2013by,Boyanovsky:2015tba,Guilleux:2015pma,Moss:2016uix,Prokopec:2017vxx} for various examples. The most prominent one is certainly the stochastic approach, developed in Ref.~\cite{Starobinsky:1994bd}. It gives an effective description of the dynamics of the infrared, long wavelength, modes in terms of an effective Langevin equation. The infrared modes of the scalar fields behave classically as a result of the aforementioned gravitational amplification and experience a random noise which encodes the effect of the ultraviolet modes crossing the horizon during expansion. The Langevin dynamics can be treated through the equivalent Fokker-Planck equation. This gives access, for example, to the late-time, equilibrium probability distribution for the fields, from which one can compute various equal-time correlators, often analytically for simple enough potentials. Unequal time correlators or genuine nonequilibrium properties, which contain important information about the long time/distance properties of the theory (dynamical timescales, spectral indices, etc.), are more difficult to access analytically and even numerically in some situations. For instance, for a simple quartic potential, the cases of vanishing or of negative square mass are intrinsically nonperturbative. 

The stochastic Langevin equation is a particular case of the so-called model A in the Halperin {\it et al.} classification of nonequilibrium dynamical systems \cite{Hohenberg:1977ym}. In the present article, we shall use tools developed in this context to compute various unequal time correlators at large time separation, which gives access to different autocorrelation and relaxation timescales. In the stationary state, the problem can be formulated as a supersymmetric one-dimensional field theory \cite{Janssen:1976,Canet:2011wf,Prokopec:2017vxx}, free of ultraviolet divergences and which is a lot easier to manage than the original $D$-dimensional quantum field theory (QFT). 

This one-dimensional field theory gives analytic access to properties of the stationary state reached by the scalar fields in the late-time limit. Diagram resummations, previously performed in the complete four-dimensional field theory \cite{Gautier:2013aoa,Gautier:2015pca}, can be done here in a simpler way which reproduces the leading infrared behavior. We compute various correlators in two approximations schemes. First, in a perturbative expansion in the self-interaction coupling constant, which is, however, limited to not too light fields. The second approximation scheme is the $1/N$ expansion, where $N$ is the number of scalar fields. The latter allows us to consider the interesting case of massless fields and of a symmetry breaking potential \cite{Gautier:2015pca,LopezNacir:2019ord}. 

Along with the path integral formulation of the model A comes some interpretation of the different correlators and specific relations which are usually formulated in a statistical physics language. Using this analogy allows us here to reformulate these results in terms of our particular model and discuss some consequences for the scalar field correlator.

After briefly reviewing the effective stochastic approach, we present the functional formulation of the Langevin equation and discuss the supersymmetry of the resulting field theory in Sec.~\ref{sec:setup}. Various properties of the field correlators, independent of any approximation scheme are discussed in Sec.~\ref{sec:generalfeatures}. Our calculations in the perturbative and the $1/N$ expansions are presented in Sec.~\ref{sec:resummations}. We conclude in Sec.~\ref{sec:Concl}. Additional calculations and technical details are presented in the various appendices. 

\section{General setup}
\label{sec:setup}

We briefly recall the effective stochastic theory for the superhorizon modes of light scalar fields in de Sitter spacetime and review the functional formulation of the resulting one-dimensional model A as a supersymmetric field theory. We consider an $O(N)$-symmetric scalar field theory on the expanding Poincar\'e patch of a $D$-dimensional de Sitter spacetime with $d$ spatial dimensions ($D=d+1$). The metric reads $\dd s^2 = -\dd t^2 + a(t) \dd \vec x^2$, with $a(t)=e^{Ht}$, where $t$ is the cosmological time and we set the Hubble rate $H=1$. The classical action reads
\begin{equation}
 {\cal S}=-\int_x\left\{\frac{1}{2}\partial_\mu\hat\varphi_a\partial^\mu\hat\varphi_a+\hat V\left(\hat\varphi^2\right)\right\},
\end{equation}
where $\hat\varphi^2=\hat\varphi_a\hat\varphi_a$ and $\int_x$ denotes the appropriate, invariant integration measure.

\subsection{Effective stochastic approach}

For light fields in units of $H$, the (quantum) fluctuations of long wavelength, superhorizon modes are well described by the effective Langevin equation \cite{Starobinsky:1994bd}
\begin{equation}
    \dot{ \hat \varphi}_a+ \frac1 {d} \hat V_{,\hat a} = \hat\xi_a,
    \label{eq:stochastic1}
\end{equation}
where the dot denotes a time derivative and we used the notation $\hat V_{, \hat a}=\partial\hat V/\partial\hat \varphi_a$. Here, the infrared fields $\hat\varphi_a$, spatially smeared over a Hubble patch, effectively behave as classical stochastic fields whose fluctuations mimic those of the long wavelength modes of the original quantum fields. Those stochastic fluctuations are driven by the random kicks from the (quantum) subhorizon modes which cross the horizon at a constant rate due to the gravitational redshift. This is represented by the noise term $\hat\xi_a$, whose stochastic properties reflect the quantum state of the system. For the Bunch-Davies (BD) vacuum, and treating the ultraviolet modes in the linear approximation, one finds \cite{Starobinsky:1994bd}
\begin{equation}
    \ev{\hat\xi_a(t,\vec x) \hat\xi_b(t',\vec x')} = \frac2{d\Omega_{D+1}} \delta_{ab} \delta(t-t') {\cal F}(\abs{\vec x - \vec x'}),
    \label{eq:noisestochastic1}
\end{equation}
with $\Omega_{n} = 2\pi^{n/2}/\Gamma\qty(n/2)$ and where the function ${\cal F}$ reflects the spatial smearing: it can always be normalized as ${\cal F}(0)=1$ and it vanishes rapidly for spatial separations $\abs{\vec x - \vec x'}\gtrsim1$. Its precise form depends on the smearing procedure. Within a single Hubble patch, ${\cal F}\approx1$ and the time evolution of the infrared fields is described by an effective one-dimensional Langevin equation with a Gaussian white noise. At sufficiently late times, the system is driven towards a stationary regime where, {\em e.g.}, the equilibrium distribution of field values is given by
\begin{equation}
    P(\hat \varphi_a) \propto e^{-\Omega_{D+1} \hat V(\hat \varphi^2)}.
    \label{eq:equilibrium}
\end{equation}
The latter describes the equal-time statistical properties of the stochastic process and reflects the quantum fluctuations of the infrared modes of the original quantum fields in the BD vacuum. It can be seen as the Boltzmann distribution for a thermal system. Introducing the Hamiltonian for the superhorizon field in the Hubble patch under consideration as $\hat {\cal H}=\int d^dx \hat V={\cal V}_d \hat V$, where ${\cal V}_d=\Omega_d/d$ is the volume of the $d$-dimensional spherical Hubble patch (of radius $H^{-1}=1$), the distribution \eqref{eq:equilibrium} reads ${\cal P}\propto e^{-\beta \hat{\cal H}}$, 
with $\beta=\Omega_{D+1}/{\cal V}_d=2\pi$ the inverse Gibbons-Hawking temperature \cite{Gibbons:1977mu}.

It is useful to rescale the variables so as to absorb the various volume factors. Defining 
\begin{align}\label{eq:rescaling}
 \hat\varphi_a = \sqrt{\frac2{d\Omega_{D+1}}}\varphi_a \quad{\rm and}\quad\hat V(\hat\varphi^2) = \frac2{\Omega_{D+1}}V(\varphi^2),
\end{align}
we get, denoting the time derivative with a dot,
\begin{equation}
     \dot\varphi_a(t) + V_{,a}(t) = \xi_a(t)
    \label{eq:staro}
\end{equation}
\begin{equation}
    \ev{\xi_a(t) \xi_b(t')} = \delta_{ab} \delta(t-t')
    \label{eq:whitenoise}
\end{equation}
This is a particular, $(0+1)$-dimensional case of the model A in the classification of Halperin {\it et al.} \cite{Hohenberg:1977ym}, which has been widely studied in the context of out-of-equilibrium statistical physics. It can be given an elegant functional formulation by means of the Janssen-de Dominicis (JdD) procedure \cite{Janssen:1976,Canet:2011wf}, which provides an efficient starting point for implementing various field techniques \cite{Canet:2011wf}. Recent examples in the present context include diagrammatic methods \cite{Garbrecht:2013coa,Garbrecht:2014dca} or renormalization group techniques \cite{Prokopec:2017vxx}. We now briefly review the JdD procedure.

\subsection{Path integral formulation}

The expectation value of an operator $\mathcal{O}(\varphi)$ can be formally expressed as
\begin{equation}
    \ev{\mathcal{O}(\varphi)} = \int \mathcal{D} \xi P[\xi] \mathcal{O}(\varphi_{\xi}) 
    \label{eq:JDD}
\end{equation}
where $\varphi_{\xi}$ is a solution of Eq.~\eqref{eq:staro} with given initial conditions and 
\begin{equation}
P[\xi] = \frac1{\sqrt{2 \pi}} e^{- \int_t \frac12\xi^2}
\end{equation}
is the normalized probability distribution of the noise, with $\int_t=\int_{-\infty}^{+\infty} \dd{t}$. 
In general, one should also average over initial conditions in Eq.~\eqref{eq:JDD}. However, the latter becomes irrelevant if we restrict our considerations to the stationary regime. Assuming the uniqueness of the solution of Eq.~\eqref{eq:staro} for a given realization of the noise (and given initial conditions), one writes 
\begin{equation}
    \mathcal{O}(\varphi_{\xi}) = \int \mathcal{D}\varphi\; \delta[\dot\varphi_a + V_{,a} - \xi_a] \mathcal{J}[\varphi] \mathcal{O}(\varphi) 
    \label{eq:avO}
\end{equation}
where $\mathcal{J}[\varphi] = \abs{{\rm Det}\qty[ \delta_{ab} \partial_t + V_{,ab} ]}$ is the appropriate functional Jacobian. Under the above uniqueness assumption, one can forget the absolute value on the determinant and exponentiate the latter in terms of Grassmann fields
\begin{equation}
    \mathcal{J}[\varphi] \to \int \mathcal{D}[\psi,\bar \psi] \,e^{i\int_t  \bar \psi_a(\delta_{ab} \partial_t + V_{,ab})\psi_b }.
    \label{}
\end{equation}
Similarly, one exponentiates the functional delta as 
\begin{equation}
    \delta[\dot \varphi_a + V_{,a} - \xi_a] = \int \mathcal{D}[i\tilde \varphi] e^{-\int_t \tilde \varphi_a (\dot\varphi_a + V_{,a} - \xi_a )} ,
    \label{}
\end{equation}
where the so-called response fields $\tilde \varphi_a$ are purely imaginary. 
Integration over the Gaussian noise $\xi_a$ finally gives, up to an irrelevant constant factor ${\cal N}$,
\begin{equation}
    \ev{\mathcal{O}(\varphi)} = {\cal N}\int \mathcal{D}[\varphi,i\tilde \varphi,\psi,\bar\psi]  \; e^{-S_{\rm JdD}[\varphi,\tilde \varphi,\psi,\bar \psi]} \mathcal{O}(\varphi) ,
    \label{eq:JDDpathinteg}
\end{equation}
with the following action 
\begin{equation}
        S_{\rm JdD} =  \int_t\left\{ \tilde \varphi_a \qty( \dot \varphi_a + V_{,a} ) - \frac12 \tilde \varphi^2  -i\bar \psi_a\qty(\delta_{ab} \partial_t + V_{,ab})\psi_b\right\}.
    \label{eq:action}
\end{equation}
This one-dimensional statistical field theory with $4N$ fields describes the leading infrared behavior of the underlying QFT in de Sitter spacetime.  Alternatively, we can use a more symmetric form of the action by changing the variable $\tilde{\varphi}_a\to F_a = i (\dot \varphi_a - \tilde \varphi_a)$. The action rewrites as
    \begin{align}
    \label{eq:actionF}
        S_{\rm JdD} = & \int_t \left\{\frac12 \dot \varphi^2 + \frac12 F^2 - i \bar \psi_a \dot\psi_a+ iF_aV_{,a} - i \bar\psi_a V_{,ab} \psi_b \right\},
    \end{align}
where we neglect the boundary term $\int_t 2\dot\varphi_aV_a=\int_t\dot V$ in the stationary state. This form of the action makes clear another link, namely, it relates to a supersymmetric quantum mechanics after the Wick rotation $t\to i\tau$ \cite{Synatschke:2008pv}.

\subsection{Supersymmetry}

The action \eqref{eq:action} or, equivalently, \eqref{eq:actionF}, possesses various symmetries, such as  the time-translation and the time-reversal symmetries of the stationary regime, which can be conveniently encoded in a supersymmetry that mixes the bosonic and fermionic degrees of freedom \cite{Canet:2011wf,Synatschke:2008pv}.  To exhibit the latter, it is convenient to recast the various fields into the superfield 
\begin{equation}\label{eq:superfield}
\Phi_a(t,\theta,\bar\theta) = \varphi_a(t) + \bar \theta \psi_a(t) + \bar \psi_a(t) \theta + \bar \theta \theta F_a(t),
\end{equation}
 living on the superspace $(t,\theta,\bar\theta)$, with Grassmann directions $\theta$ and $\bar \theta$. The generators of the supersymmetry can be written as $Q = i\partial_{\bar \theta} + \theta\partial_t$ and $\bar Q = i\partial_\theta + \bar \theta \partial_t$, and the covariant derivatives $D=i \partial_{\bar \theta} - \theta \partial_t$, $\bar D = i\partial_\theta - \bar \theta \partial_t$ allow us to write the action in the following form 
\begin{equation}
    S_{\rm JdD} = \int \dd{z} \left\{\frac12 \Phi_a K \Phi_a + i V(\Phi_a)\right\} 
    \label{}
\end{equation}
with\footnote{Our convention for the Grassmann integration is $\int \dd\theta \dd\bar\theta \bar \theta\theta = 1$.} $z = (t,\bar\theta,\theta)$, $\dd{z} = \dd{t}\dd\theta\dd\bar\theta$ and $K=\frac12 \qty(\bar D D - D\bar D)$. 

\section{General properties of the correlator}
\label{sec:generalfeatures}

The general form of the superfield correlators is constrained by various considerations, most prominently the symmetries and causality. In this section, we detail the case of the connected\footnote{Unless explicitly stated, we only consider connected correlators in what follows. For simplicity, we do not introduce a special notation.} two-point correlator, with the notation
\begin{equation}
 G^{ab}_{12}(t_1,t_2)=\ev{\Phi_a(t_1,\theta_1,\bar\theta_1)\Phi_b(t_2,\theta_2,\bar\theta_2)}.
\end{equation}
For simplicity, we consider a single field ($N=1$). The generalization to arbitrary $N$ is trivial.

\subsection{Supersymmetry constraints}

The dependence of the inverse propagator $\Gamma^{(2)}$ and the propagator $G$ on the Grassmann variables is strongly constrained by the supersymmetry of the action. First, the anticommutator $\acomm{Q}{\bar Q}=2i\partial_t$ generates the time-translation invariance, so that it proves more convenient to work in frequency space
\begin{equation}
 G_{12}(t_1,t_2)=\int\frac{\dd{\omega}}{2\pi} e^{-i\omega (t_1-t_2)}G_{12}(\omega).
\end{equation}
and similarly for the two-point vertex $\Gamma^{(2)}_{12}(\omega)$. The general dependence of the latter in the Grassmann variables involves {\it a priori} six independent functions:
\begin{align}
        \Gamma^{(2)}_{12}(\omega) &= A(\omega) + \bar \theta_1 \theta_1 B(\omega) + \bar \theta_2 \theta_2 C(\omega) + \bar\theta_1 \theta_1 \bar \theta_2 \theta_2 D(\omega)\nonumber \\
        & + \bar \theta_1 \theta_2 E(\omega) + \bar \theta_2 \theta_1 F(\omega) .
    \label{}
\end{align}
Supersymmetry implies the Ward identities
\begin{align}
 \qty(Q_1 + Q_2)\Gamma^{(2)}_{12}(\omega) &= 0,\\
 \qty(\bar Q_1 + \bar Q_2)\Gamma^{(2)}_{12}(\omega) &= 0,
\end{align}
where the numerical index indicates the Grassmann variable each operator $Q$ or $\bar Q$ is acting on. These yield four independent constraints which are solved as
\begin{align}
 C(\omega)&=B(\omega)\\
 D(\omega)&=\omega^2A(\omega)\\
 E(\omega)&=-B(\omega)-\omega A(\omega)\\
 F(\omega)&=-B(\omega)+\omega A(\omega).
\end{align}
Renaming $A(\omega)=\eta(\omega)$ and $B(\omega)=i\gamma(\omega)$, the general structure of the two-point vertex is  \cite{Canet:2011wf,Synatschke:2008pv}
\begin{equation}
    \Gamma^{(2)}_{12}(\omega) = i\gamma(\omega)\delta_{12} + \eta(\omega) K_{\omega}\delta_{12},
    \label{eq:gamma2general}
\end{equation}
where the two Grassmann structures
\begin{align}
 \delta_{12} &= (\bar \theta_1 - \bar \theta_2)(\theta_1 - \theta_2),\\
 K_\omega\delta_{12} &= 1 + \omega (\bar \theta_2 \theta_1 - \bar \theta_1 \theta_2) + \omega^2\bar \theta_1 \theta_1 \bar \theta_2 \theta_2 
\end{align}
denote, respectively, the Dirac function in Grassmann coordinates and the supersymmetric d'Alembertian operator $K_1\delta(z_1-z_2)$ in frequency space, with $\delta(z_1-z_2)=\delta(t_1-t_2)\delta_{12}$. 

The superfield propagator is obtained by inversion, $\int_2 \Gamma^{(2)}_{12}(\omega)G_{23}(\omega)=\delta_{13}$, with $\int_2=\int\dd{\theta_2}\dd{\bar\theta_2}$, and reads
\begin{align}
     G_{12}(\omega)= \frac{-i\gamma(\omega) \delta_{12} + \eta (\omega)K_\omega\delta_{12}}{\omega^2 \eta^2(\omega) + \gamma^2(\omega)}.
    \label{eq:propaggeneral}
\end{align}
Using the decomposition \eqref{eq:superfield} of the superfield, we obtain the various correlators\footnote{Our convention is $\ev{A(t)B(t')}=\int\frac{\dd{\omega}}{2\pi}e^{-i\omega(t-t')}G_{AB}(\omega)$.}
\begin{align}
\label{eq:Gphiphi}
G_{\varphi\varphi}(\omega) &= \frac{\eta(\omega)}{\omega^2 \eta^2(\omega) + \gamma^2(\omega)} ,\\
G_{\varphi F}(\omega) &=  \frac{-i\gamma(\omega)}{\omega^2 \eta^2(\omega) + \gamma^2(\omega)}, 
\end{align}
as well as $G_{FF}(\omega)=\omega^2G_{\varphi\varphi}(\omega)$, $G_{\psi\bar\psi}(\omega)=-G_{\varphi F}(\omega) - \omega G_{\varphi\varphi}(\omega)$, and $G_{\bar\psi\psi}(\omega)=G_{\varphi F}(\omega) - \omega G_{\varphi\varphi}(\omega)$. 
Now, from the path integral representation \eqref{eq:JDDpathinteg}, we see that both $\ev{\varphi(t)\varphi(t')}$ and $\ev{\varphi(t)\tilde\varphi(t')}$ are real (despite $\tilde\varphi$ being  imaginary) and thus $G_{\varphi\varphi}(t)\in \mathbb{R}$ and $G_{\varphi F}(t)\in i\mathbb{R}$. Using also the permutation identity of the superfield correlator, $G_{12}(t)=G_{21}(-t)$, we conclude, in frequency space, that both the functions $\gamma(\omega)$ and $\eta(\omega)$ are real and even. 

\subsection{Fluctuation-dissipation relation}

The stationary,  equilibrium state of the system is characterized by a fluctuation-dissipation relation which directly follows from the above constraints. This relates the statistical correlator $G_{\varphi\varphi}(\omega)$ (fluctuation) to the response function\footnote{The relation of the stochastic response and spectral functions with the retarded and spectral functions of the underlying QFT are discussed in the Appendix~\ref{sec:quantumvsstochastic}.} $G_{\varphi\tilde\varphi}(\omega)$ (dissipation) or, more precisely, to the stochastic spectral function $\rho$, which we now introduce. The response function is given by
\begin{align}
    G_{\varphi\tilde\varphi}(\omega)= i\qty[G_{\varphi F}(\omega) + \omega G_{\varphi\varphi}(\omega)]= \frac{i}{\omega \eta(\omega) + i\gamma(\omega)}
    \label{eq:Gretarded}
\end{align}
and we define the stochastic spectral function as 
\begin{equation}
    \rho(\omega) \equiv 2 i \Im G_{\varphi\tilde\varphi}(\omega)=2i\omega G_{\varphi\varphi}(\omega),
    \label{eq:spectral}
\end{equation} 
where the second equality follows from Eqs.~\eqref{eq:Gphiphi}--\eqref{eq:Gretarded}. In real time, this reads
\begin{equation}
\label{eq:fdrealtime}
  \rho(t)=-2 \partial_t G_{\varphi\varphi}(t).
\end{equation} 
This is the announced fluctuation-dissipation relation characteristic of a thermal state in the high temperature (classical field) regime as discussed in the Appendix~\ref{sec:quantumvsstochastic}. 

An interesting consequence of the above relation is the exact identity
\begin{equation}\label{eq:rhoone}
\rho(t=0^+)= 1,
\end{equation}
which can be proven as follows. In the limit\footnote{The correlator $\ev{\xi(0^+)\varphi(0)}=0$ by causality. Considering, instead, $t\to0^-$, one would have to take into account the nonzero correlator $\ev{\xi(0^-)\varphi(0)}$. The final result is $\partial_t G_{\varphi\varphi}(t)|_{t\to0^-}=-\partial_t G_{\varphi\varphi}(t)|_{t\to0^+}=1/2$.} $t\to0^+$, we have, using the relation \eqref{eq:fdrealtime} and Eq.~\eqref{eq:staro}
\begin{align}
 \rho(t=0^+)=-2\partial_t G_{\varphi\varphi}(t)|_{t\to0^+}=-2\ev{\dot\varphi\varphi}=\ev{2\varphi \partial_\varphi V}.
\end{align}
The equal-time average in the last equality can be computed with the one-point equilibrium distribution \eqref{eq:equilibrium} with the proper rescaling \eqref{eq:rescaling}. The result \eqref{eq:rhoone} follows from the identity
\begin{align}
 \int_{-\infty}^{+\infty}\dd{\varphi}2\varphi\qty(\partial_\varphi V)e^{-2V}=\int_{-\infty}^{+\infty}\dd{\varphi}e^{-2V},
\end{align}
obtained after integration by parts.

\subsection{Causality}

Further interesting information can be obtained from causality. The latter implies, in particular, that the response function vanishes identically for negative times, $G_{\varphi\tilde\varphi}(t)\propto \theta(t)$ \cite{Canet:2011wf}. From the definition \eqref{eq:spectral} of the spectral function and the fact that $G_{\varphi\tilde\varphi}(t)\in\mathds{R}$, we easily deduce that $\rho(t)=G_{\varphi\tilde\varphi}(t)-G_{\varphi\tilde\varphi}(-t)$ and thus that\footnote{This is equivalent to $G_{\varphi F}(t) = i \,{\rm sign}(t)\, \partial_t G_{\varphi\varphi}(t)$ \cite{Zinn-Justin:1996} }
\begin{equation}
 G_{\varphi\tilde\varphi}(t)=\theta(t)\rho(t),
\end{equation}
or, equivalently, in frequency space, 
\begin{equation}
 G_{\varphi\tilde\varphi}(\omega)=\int\frac{\dd{\omega'}}{2\pi}\frac{i\rho(\omega')}{\omega-\omega'+i0^+},
\end{equation}
which implies that $G_{\varphi\tilde\varphi}$ is analytic in the upper half complex frequency plane. 

Also, using the fluctuation-dissipation relation \eqref{eq:spectral}, we deduce
\begin{equation}
    G_{\varphi\tilde\varphi}(\omega=0)= 2 \int \frac{\dd{\omega}}{2\pi}G_{\varphi\varphi}(\omega)=2G_{\varphi\varphi}(t=0).
    \label{eq:ggrel}
\end{equation}
This yields an exact expression for the so-called\footnote{Note though that this is actually a static (equal-time) quantity.} dynamical mass $m_{\rm dyn}$, which measures the amplitude of the equal-time fluctuations of the stochastic field within a Hubble patch as 
\begin{equation}\label{eq:mdyn}
 G_{\varphi\varphi}(t=0)=\ev{\varphi^2} \equiv \frac{1}{2m_{\rm dyn}^2}.
\end{equation}
Using Eqs.\eqref{eq:Gretarded} and \eqref{eq:ggrel}, we deduce 
\begin{equation}
    m_{\rm dyn}^2 = \gamma(0).
    \label{eq:mdyngamma}
\end{equation}
Such a relation is reminiscent of the concept of screening mass, or susceptibility in thermal (quantum/statistical) field theory, which are related to the value of the (inverse) propagator at vanishing momentum and frequency and typically measure the overall response of the system to a static perturbation. These are to be distinguished from the so-called pole masses, or correlation lengths, which are associated to the poles of the response function and describe correlations between different spacetime points. The latter have their analogs in the present stochastic model, which we now discuss.

\subsection{Mass hierarchy} 

Using the Fokker-Planck formulation of the Langevin equation \eqref{eq:staro}, one shows that the unequal time (connected) correlator for a given local function ${\cal A}(\varphi)$ of the field can be written as \cite{Starobinsky:1994bd,Markkanen:2019kpv} 
\begin{equation}
  G_{{\cal A}{\cal A}}(t-t') = \ev{{\cal A}(t){\cal A}(t')}=\sum_{n\ge0} \sum_{\ell=0}^nC^{\cal A}_{n,\ell} e^{-\Lambda_{n,\ell} |t-t'|},
    \label{eq:AAdecomp}
\end{equation}
where the $\Lambda_{n,\ell}$'s are the eigenvalues of the (properly rescaled) Fokker-Planck operator and the $C^{\cal A}_{n,\ell}$'s are appropriate coefficients.
Because of the O($N$) symmetry, the latter can be labeled in terms of the eigenvalues $\ell\in\mathds{N}$ of the $N$-dimensional angular momentum and another possible index $n$. In the case of a quadratic potential, the latter is a single positive integer and the possible values of $\ell$ are constrained such that $n-\ell$ is even and positive. We expect this to remain true for $\lambda\neq0$.

The eigenvalues are non-negative real numbers\footnote{Supersymmetry guarantees that the lowest eigenvalue $\Lambda_{0,0}=0$.}. Of course, some $C_{n,\ell}^{\cal A}$ may vanish, {\it e.g.}, due to symmetry selection rules \cite{Markkanen:2019kpv}. For instance, the case ${\cal A}=\varphi$ only involves the vector channel $\ell=1$, so that the only nonvanishing coefficients in the decomposition \eqref{eq:AAdecomp} are $C^\varphi_{2n+1,1}$. Similarly, for the composite field $\chi=\varphi^2/(2N)$ in the scalar ($\ell=0$) channel, the only nonvanishing terms are $C_{2n,0}^\chi$. The correlations of various quantities of interest at large time separations are thus governed  by the lowest eigenvalues contributing to the decomposition \eqref{eq:AAdecomp}.

Below, we shall compute the $\ev{\varphi\varphi}$ and  $\ev{\chi\chi}$ correlators in various approximation schemes, from which we can extract the eigenvalues $\Lambda_{2n+1,1}$ and $\Lambda_{2n,0}$, respectively, at each approximation order. Introducing the redefinitions $C^\varphi_{2n+1,1}=c^\varphi_{2n+1}/(2\Lambda_{2n+1,1})$, $C^\chi_{2n,0}=c^\chi_{2n}/(2\Lambda_{2n,0})$, and the following notation for the tree-level correlator of a field of mass $m$
\begin{equation}\label{eq:freeprop}
 G_{m^2}(t)=\frac{e^{-m^2|t|}}{2m^2}\quad\Leftrightarrow\quad G_{m^2}(\omega)=\frac{1}{\omega^2+m^4},
\end{equation}
we have
\begin{align}
    \label{eq:propagphi}
    G_{\varphi\varphi}(t) &=\sum_{n\ge0} c^\varphi_{2n+1}G_{\Lambda_{2n+1,1}}(t),\\
    \label{eq:propagchi}
    G_{\chi\chi}(t) &= \sum_{n\ge0} c^\chi_{2n}G_{\Lambda_{2n,0}}(t).
\end{align}
The eigenvalues $\Lambda_{n,\ell}$ and the coefficients $c^{\varphi,\chi}_{n}$ are directly obtained as the poles and residues of the relevant response function, {\it e.g.},
\begin{equation}\label{eq:responsefraction}
G_{\varphi\tilde\varphi}(\omega)=\sum_{n\ge0}  \frac{ic^\varphi_{2n+1}}{\omega+i\Lambda_{2n+1,1}}.
\end{equation}

An obvious relation is 
 \begin{equation}
     \sum_{n\ge0} \frac{c^\varphi_{2n+1}}{\Lambda_{2n+1,1}}=\frac1{m_{\rm dyn}^2},
    \label{eq:mdyn2}
\end{equation}
which directly follows from the definition \eqref{eq:mdyn}. Another constraint on the coefficients $c^\varphi_{2n+1}$ is the 
 following sum rule
\begin{equation}\label{eq:sumrule}
\sum_{n\ge0}c^\varphi_{2n+1}=-2\partial_t G_{\varphi\varphi}(t)|_{t\to0^+}= 1,
\end{equation}
which directly follows from Eqs.~\eqref{eq:fdrealtime} and \eqref{eq:rhoone}.

\subsection{Effective noise correlator}
\label{sec:noisecorrelator}

Finally, we mention that the $\eta$ component of the self-energy \eqref{eq:gamma2general} can be interpreted as the effective noise correlator dressed by the nonlinear effect of the infrared modes themselves. Indeed, as recalled in the Appendix~\ref{sec:quantumvsstochastic}, the general expression of the correlator of a Langevin process with a colored noise 
\begin{align}
 \ev{\xi(t)\xi(t')} =\int\frac{\dd\omega}{2\pi}e^{-i\omega(t-t')}{\cal N}(\omega)
\end{align}
is, in frequency space,
\begin{equation}
 G_{\varphi\varphi}(\omega)={\cal N}(\omega)|G_{\varphi\tilde\varphi}(\omega)|^2.
\end{equation}
Using the exact relations \eqref{eq:Gphiphi} and \eqref{eq:Gretarded}, we deduce that
\begin{equation}\label{eq:noiseandsigma}
 {\cal N}(\omega)=\eta(\omega)
\end{equation}
can be interpreted as an effective colored noise kernel as announced. The tree-level expression $\eta_{\rm free}(\omega)=1$ corresponds to the white noise contribution \eqref{eq:whitenoise} from the ultraviolet modes in the present effective stochastic theory. As we shall see below, nonlocal loop corrections bring a nontrivial frequency dependence which corresponds to the effective dressing of the noise kernel from the nonlinear infrared dynamics.

\section{Explicit calculations}
\label{sec:resummations}

\begin{figure}[t]
    \centering
    \includegraphics{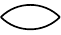}
    \caption{One-loop diagram giving the expression of $C_{12}^{m^2}$ in a free theory. The lines denote the tree-level propagator \eqref{eq:treeprop}.}
    \label{fig:onebubble}
\end{figure}

We now turn to explicit computations of the $\ev{\varphi\varphi}$ and  $\ev{\chi\chi}$ correlators in two approximation schemes previously studied in the D-dimensional QFT \cite{Gautier:2013aoa,Gautier:2015pca}, namely, the perturbative expansion and the $1/N$ expansion. 
We consider an $O(N)$-symmetric scalar theory with quartic self-interaction, whose superpotential is given by
\begin{equation}
    V(\Phi) = \frac{m^2}2 \Phi_a^2 + \frac\lambda{4!N} \qty(\Phi_a^2)^2 .
    \label{eq:potential}
\end{equation}
There is no possibility of spontaneously broken symmetry in the present low dimensional system \cite{Mermin:1966fe,Coleman:1974jh,Serreau:2013eoa}. We thus have $\ev{\Phi_a}=0$ and $G_{12}^{ab}(\omega)=G_{12}(\omega) \delta^{ab}$, including in the case $m^2<0$. 

In the following, we define the superfield self-energy $\Sigma$ as 
\begin{equation}
    \Gamma^{(2)}_{12}(\omega) =im^2 \delta_{12} + K_\omega\delta_{12} + \Sigma_{12}(\omega)
    \label{eq:}
\end{equation}
where the first two terms on the right-hand side correspond to the free-field case. We denote the tree-level superpropagator for a field with mass $m$ as
\begin{equation}\label{eq:treeprop}
 G_{12}^{m^2}(\omega)=\frac{-im^2\delta_{12}+K_\omega\delta_{12}}{\omega^2+m^4}.
\end{equation}
We also introduce the supercorrelator of the composite field $X=\Phi^2/(2N)$,
\begin{equation}
    C_{12}(t) = \ev{X(t,\bar \theta_1,\theta_1)X(0,\bar\theta_2,\theta_2)},
    \label{eq:phi2phi2}
\end{equation}
which, in the free theory, is simply given by the one-loop diagram of Fig.~\ref{fig:onebubble}. This is easily computed as [see Eq.~\eqref{eq:relation2}] 
\begin{align}
        C^{m^2}_{12}(\omega) &= \frac1{2N} \int\frac{\dd{\omega'}}{2\pi}G^{m^2}_{12}(\omega-\omega')G^{m^2}_{12}(\omega')=\frac{ G^{2m^2}_{12}(\omega)}{2Nm^2}.
    \label{eq:Cm2}
\end{align}
The component at $\theta_{1,2}=\bar\theta_{1,2}=0$ is
\begin{equation}
        G^{m^2}_{\chi\chi}(\omega) =  \frac1{2N m^2} \frac1{\omega^2+4m^4}.
    \label{eq:Gchifree}
\end{equation}
From the decompositions \eqref{eq:propagphi} and \eqref{eq:propagchi} and the free-field expressions \eqref{eq:freeprop} and \eqref{eq:Gchifree}, we read $\Lambda^{\rm free}_{1,1}=m^2$, $c^{\rm free}_{2n+1}=\delta_{n,0}$, $\Lambda^{\rm free}_{2,0}=2m^2$, and $c^{\rm free}_{2n}=\delta_{n,1}/(2Nm^2)$. This agrees with the known spectrum of the free case, which is just that of a O($N$)-symmetric harmonic oscillator \cite{Starobinsky:1994bd,Markkanen:2019kpv} 
\begin{equation}
     {\Lambda_{n,\ell}^{\rm free}} = n m^2.  
    \label{eq:lambdafree}
\end{equation}

\subsection{The perturbative expansion}
\label{sec:perturbative}

\begin{figure}[t]
    \centering
    \includegraphics[width=8cm]{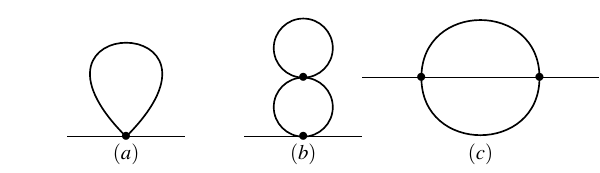}
    \caption{Perturbative contributions to the self-energy $\Sigma$ at one- and two-loop orders. The interactions vertex is represented with a dot and contributes a factor $-i\lambda/(4!N)$ while the propagator lines are given by the tree-level propagator \eqref{eq:treeprop}.}
    \label{fig:perturbative}
\end{figure}

We first compute the self-energy at two-loop order in a perturbative expansion (the three-loop order is computed in Appendix \ref{sec:perturbative4}). The relevant diagrams are shown in Fig.~\ref{fig:perturbative}. Their explicit evaluation is straightforward and we shall only give the resulting expressions here. The details can be found in Appendix  \ref{sec:perturbative3}. The one-loop contribution, diagram (a), yields
\begin{equation}
    \Sigma^{(2a)}_{12}(\omega) = i \frac{N+2}{3N} \frac\lambda{4m^2} \delta_{12},
    \label{eq:sigma1looppert}
\end{equation}
which simply corresponds to a constant shift of $\gamma(\omega)$, that is, a mere mass renormalization. 
The same is true for the two-loop local\footnote{Here, local means that both external legs are attached to the same vertex.} contribution given by diagram $(b)$ in Fig.~\ref{fig:perturbative}, which reads
\begin{equation}
    \Sigma^{(2b)}_{12}(\omega) = -i \qty(\frac{N+2}{3N})^2 \frac{\lambda^2}{16 m^6} \delta_{12}.
    \label{eq:sigma2looppert}
\end{equation}
A nontrivial frequency dependence appears with the nonlocal contribution, diagram $(c)$, which can be written as
\begin{equation}
    \Sigma^{(2c)}_{12}(\omega) = \frac{N+2}{3{ N^2}} \frac{\lambda^2}{8m^4} G^{3m^2}_{12}(\omega).
    \label{eq:selfenergy}
\end{equation}
Altogether, we obtain, for the functions $\gamma$ and $\eta$ in Eq.~\eqref{eq:gamma2general},
\begin{align}
\label{eq:gammatwoloop}
        \gamma(\omega) &= M^2 - \frac{6\bar\lambda^2}{N+2}\frac{3 m^6}{\omega^2 + 9m^4} + {\cal O}({\bar\lambda^3}) ,\\
\label{eq:etatwoloop}
        \eta(\omega) &= 1 +\frac{6\bar\lambda^2}{N+2}\frac{m^4}{\omega^2 + 9m^4} + {\cal O}({\bar\lambda^3}) ,
\end{align}
where we have introduced the dimensionless coupling
\begin{equation}\label{eq:barlambda}
 \bar\lambda=\frac{N+2}{3N}\frac\lambda{4m^4}
\end{equation}
and the renormalized mass
\begin{equation}
    M^2 = m^2 \left(1+ \bar\lambda - \bar\lambda^2\right).
    \label{eq:renormmass}
\end{equation}

We immediately obtain the expression of the dynamical mass as
\begin{equation}
 m_{\rm dyn}^2=\gamma(0)=m^2 \left[ 1 +\bar\lambda- \frac{N+4}{N+2}\bar\lambda^2 +{\cal O}({\bar\lambda^3})\right]
\end{equation}
As explained in Sec.~\ref{sec:generalfeatures}, the relevant mass hierarchy can be directly read off the response function. Using the expressions \eqref{eq:gammatwoloop} and \eqref{eq:etatwoloop}, the latter can be written as 
\begin{equation}
    G_{\varphi\tilde\varphi}(\omega) = \frac{ic_1}{\omega +i \Lambda_{1,1}} + \frac{ic_3}{\omega+i\Lambda_{3,1}}+ \order{\bar\lambda^3},
    \label{eq:propagFouriertwoloop}
\end{equation}
with the poles given by
\begin{align}
        {\Lambda_{1,1}} &= m^2 \left[ 1 +\bar\lambda- \frac{N+5}{N+2}\bar\lambda^2 +{\cal O}({\bar\lambda^3})\right],\\
        {\Lambda_{3,1}} &= 3m^2 \left[1+ {\cal O}(\bar \lambda)\right],
\end{align}
and the residues
\begin{align}
        c^\varphi_1 &= 1 - \frac{3\bar\lambda^2}{2(N+2)}+{\cal O}(\bar\lambda^3),\\
        c^\varphi_3 &=\frac{3\bar\lambda^2}{2(N+2)}+{\cal O}(\bar\lambda^3).
\end{align}
In particular, we verify the sum rule \eqref{eq:sumrule} at this order.

The two-pole structure \eqref{eq:propagFouriertwoloop} at the present order of approximation precisely coincides to the splitting of the propagator obtained in the QFT calculation of Ref.~\cite{Gautier:2013aoa}, which reads
\begin{equation}
    G_{\varphi\varphi}(t) = c_+ G_{m_+^2}(t) + c_- G_{m_-^2}(t),
    \label{eq:propagphiphi}
\end{equation}
with $G_{m^2}$ given in Eq.~\eqref{eq:freeprop}.
The expressions of the various masses and coefficients exactly agree, with the identifications $c_+=c^\varphi_1$, $c_-=c^\varphi_3$, $m_+^2= \Lambda_{1,1}$, $m_-^2= \Lambda_{3,1}$, and with the rescaling \eqref{eq:rescaling}, that is,\footnote{In particular, the quantity named $\bar\lambda$ in Ref.~\cite{Gautier:2013aoa} is the same as here.}
\begin{equation}
    m^2 = \frac{\hat m^2}d \qc \lambda  = \frac2{d^2\Omega_{D+1} }\hat\lambda.
    \label{}
\end{equation}

\begin{figure}[t]
    \centering
    \includegraphics{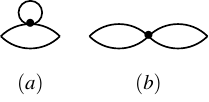}
    \caption{Two-loop contributions to the $\ev{\chi\chi}$ correlator. The diagram (a) is just an effect of the mass renormalization.}
    \label{fig:twobubbles}
\end{figure}

We now come to the two-loop correction to the $\ev{\chi\chi}$ correlator, given by the two diagrams in Fig.~\ref{fig:twobubbles}. The diagram (a) simply corresponds to the effect of the one-loop mass renormalization of one propagator line [the same is true for the diagram (b) of Fig.~\ref{fig:perturbative}] and can be easily computed. Equivalently, we can treat this diagram with the following trick \cite{Gautier:2013aoa,Gautier:2015pca}. We implicitly include it in the one-loop diagram of Fig.~\ref{fig:onebubble} by using effective propagator lines with an effective mass $M$. We then replace the latter by its expression \eqref{eq:renormmass} and systematically expand at the relevant order of approximation. 

Each loop in the diagram of \ref{fig:onebubble} and the diagram (b) of \ref{fig:twobubbles} is given by Eq.~\eqref{eq:Cm2}, with $m^2\to M^2$ and the sum reads 
\begin{align}
        C^{(1+3b)}_{12}(\omega) &=C^{M^2}_{12}\!(\omega)-i\lambda \frac{N+2}{3} \int_3C^{M^2}_{13}\!(\omega) \,C^{M^2}_{32} \!(\omega).
    \label{}
\end{align}
with $\int_{3}=\int\dd{\theta_3}\dd{\bar\theta_3}$. Using the identity
\begin{equation}
 \int_3G^{m^2}_{13}(\omega) G^{m^2}_{32} (\omega)=\frac{(\omega^2-m^4)\delta_{12}-2im^2K_\omega\delta_{12}}{(\omega^2+m^4)^2}
\end{equation}
and extracting the component at vanishing Grassmann variables, we obtain, in terms of the renormalized mass $M^2$
\begin{equation}
    \begin{aligned}
        G_{\chi\chi}(\omega) &= \frac1{2N M^2} \frac1{\omega^2+4M^4} \qty[ 1 - \frac{8\bar\lambda M^4}{\omega^2+4M^4}  + \order{\bar\lambda^2}]\\
        &= \frac1{2N M^2} \frac1{\omega^2 + 4M^4\qty(1+\bar\lambda)^2} + \order{\bar\lambda^2}.
    \end{aligned}
    \label{}
\end{equation}
In the last equation, we have used the knowledge of the general structure \eqref{eq:AAdecomp} of the correlator to resum the two-loop correction to the propagator in the appropriate form ({\it i.e.}, a correction to the corresponding self-energy). We can directly read off the expressions 
\begin{align}
  \Lambda_{2,0}&=2M^2\qty[1+\bar\lambda+{\cal O}(\bar\lambda^2)]=2m^2\qty[1+2\bar\lambda+{\cal O}(\bar\lambda^2)],\\
 c^\chi_2&=\frac{1}{2NM^2}\qty[1+{\cal O}(\bar\lambda^2)]=\frac{1}{2Nm^2}\qty[1-\bar\lambda+{\cal O}(\bar\lambda^2)].
\end{align}

We note that the perturbative calculation of the propagator at order $\bar\lambda^2$ only gives access to the leading-order (LO) expression of the infrared subleading eigenvalue $\Lambda_{3,1}$ because the corresponding coefficient $c^\varphi_3$ is, itself, of order $\bar\lambda^2$. It is interesting to push our perturbative calculation to three-loop order so as to obtain the first correction to $\Lambda_{3,1}$ and compare to the perturbative results of Ref.~\cite{Markkanen:2019kpv} obtained by directly solving the Fokker-Planck equation. We present this calculation in the Appendix \ref{sec:perturbative4}. The three-loop expressions of $m_{\rm dyn}^2$, $\Lambda_{1,1}$, $c^\varphi_1$ and $c^\varphi_3$ can be found there. Here, we simply gather the next-to-leading results for the lowest eigenvalues: 
\begin{align}
  \Lambda_{1,1} &= m^2\qty[1 +\bar\lambda+\order{\bar\lambda^2}],\\
  \Lambda_{2,0} &=2m^2\qty[1+2\bar\lambda+\order{\bar\lambda^2}],\\
  \Lambda_{3,1} &=3m^2\qty[1+\frac{5N+22}{3(N+2)}\bar\lambda+\order{\bar\lambda^2}],
\end{align}
which reproduce (and generalize to arbitrary $D$ and $N$) the perturbative results of Ref.~\cite{Markkanen:2019kpv} for $D=4$ and $N=1$ (in that case, $\Lambda_{n,\ell}=\Lambda_n$).

\begin{figure}[t]
    \centering
    \includegraphics{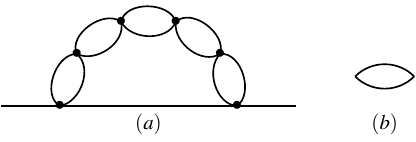}
    \caption{(a) The topology of diagrams contributing the self-energy at NLO in the $1/N$ expansion. (b) The single bubble $\Pi_{12}$.}
    \label{fig:largeN}
\end{figure}

The present perturbative calculations are controlled by the dimensionless expansion parameter $\bar\lambda\propto\lambda/m^4$ and are thus invalid in the zero mass limit as well as in the negative square mass case. These cases require a nonperturbative treatment, such as the $1/N$ expansion, studied in Ref.~\cite{Gautier:2015pca} in the QFT context and that we now describe in the present stochastic framework.

\subsection{The $1/N$ expansion}
\label{sec:largeN}

 We closely follow Ref.~\cite{Gautier:2015pca} for the diagrammatic formulation of the $1/N$ expansion, which we adapt to the present (supersymmetric) theory. In particular, we separate the local and nonlocal contributions\footnote{As mentioned earlier, local contributions consist of all diagrams where the two external legs are attached to the same vertex. These give the constant, frequency-independent contribution $\sigma$ to the function $\gamma(\omega)$ in Eq.~\eqref{eq:gamma2general}.} to the self-energy $\Sigma$ and grab the former in an effective square mass $M^2$, which satisfies the following exact gap equation:
\begin{equation}
    M^2 = m^2 + \sigma,
    \label{eq:gap}
\end{equation}
where $\sigma$ is given by the diagram $(a)$ of Fig.~\ref{fig:perturbative}, but computed with the full propagator, namely,
\begin{equation}
    \sigma = \frac{N+2}{3N} \frac\lambda2 \int\frac{\dd\omega}{2\pi} G_{11}(\omega) = \frac{N+2}{3N} \frac\lambda{4\gamma(0)} .
    \label{}
\end{equation}
Here, we have used $G_{11}(\omega)=G_{\varphi\varphi}(\omega)$ together with Eqs.~\eqref{eq:Gretarded} and \eqref{eq:ggrel}. 

In the spirit of the $1/N$ expansion, we write $M^2=M_0^2+\order{1/N}$. At LO, there are no nonlocal contributions to the self-energy and the propagator is simply given by a tree-level-like propagator $G_f^{M_0^2}$ with the LO effective mass $M_0$. In particular, we have $\gamma(0)=M_0^2+\order{1/N}$. The LO gap equation  \eqref{eq:gap} is thus solved as
\begin{equation}
M_0^2 = \frac{m^2}2 + \sqrt{\frac{m^4}{4} + \frac{\lambda}{12}}.
\end{equation}

\begin{figure}[t]
    \centering
    \includegraphics{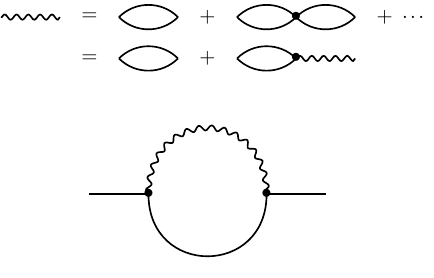}
    \caption{  Top: diagrammatic representation of the function $\mathbb{I}_{12}$ which sums the infinite series of bubble diagrams. Bottom: The nonlocal contribution to the self-energy at NLO in the $1/N$ expansion. }
    \label{fig:sumbubbles}
\end{figure}

To compute the next-to-leading (NLO) propagator, we first compute the nonlocal contributions to the self-energy $\Sigma$ at NLO in terms of the LO propagator $G_f^{M_0^2}$ (this automatically resums all LO local insertions on internal lines) and then we solve the implicit equation \eqref{eq:gap} for the local contributions at NLO. 
The NLO nonlocal contributions $\Sigma^{\rm nonloc}$ are given by the infinite series of bubble diagrams with the topology depicted in Fig.~\ref{fig:largeN}$(a)$. Each one-loop bubble, corresponding to the diagram $(b)$, gives a contribution
\begin{equation}
    \Pi_{12}(\omega) = -\frac\lambda6 \int\frac{\dd{\omega'}}{2\pi} G^{M_0^2}_{12}(\omega') G^{M_0^2}_{12}(\omega-\omega')
    \label{eq:Pi}
\end{equation}
and summing the infinite sum of bubbles is achieved by solving the integral equation
\begin{equation}
    \mathbb{I}_{12}(\omega)=\Pi_{12}(\omega) + i \int_{3} \Pi_{13}(\omega) \mathbb{I}_{32}(\omega) .
    \label{eq:integeq}
\end{equation}
The function $\mathbb{I}$ resums the infinite chain of bubble diagrams, as is depicted in Fig.~\ref{fig:sumbubbles}, where it is represented as a wiggly line. In terms of the latter the nonlocal contribution to the NLO self-energy is obtained as the bottom diagram of Fig.~\ref{fig:sumbubbles}, which gives the one-loop expression
\begin{equation}
    \Sigma^{\rm nonloc}_{12}(\omega) = -\frac\lambda{3N} \int \frac{\dd{\omega'}}{2\pi} G^{M_0^2}_{12}(\omega') \mathbb{I}_{12}(\omega-\omega').
    \label{}
\end{equation}
Again, we skip the details of the calculations and refer the reader to the Appendix \ref{sec:largeN2} for details. The calculation of the one-loop bubble follows the same lines as that of diagram (a) above. It can be written as 
\begin{equation}
    \Pi_{12}(\omega) = -2\tilde\lambda M_0^2G^{2M_0^2}_{12}(\omega)   
    \label{}
\end{equation} 
and we get, for the infinite series of bubbles,
\begin{equation}\label{eq:IIsol}
    \mathbb{I}_{12}(\omega) = -2\tilde\lambda M_0^2G^{2 M_0^2 (1+\tilde{\lambda})}_{12}(\omega),
\end{equation}
where we defined the effective dimensionless coupling 
\begin{equation}\label{eq:lambdatilde}
 \tilde{\lambda} = \frac\lambda{12M_0^4},
\end{equation}
which is the large-$N$ analog of $\bar\lambda$ defined in Eq.~\eqref{eq:barlambda}. The nonlocal self-energy at NLO reads
\begin{equation}
    \Sigma^{\rm nonloc}_{12}(\omega) = \frac{2M_0^4}N \frac{\tilde\lambda^2(3+2\tilde\lambda)}{1+\tilde\lambda} \,G^{M_0^2(3+2\tilde\lambda)}_{12}(\omega)
    \label{}
\end{equation}
and has a similar structure as the two-loop nonlocal self-energy in the previous perturbative calculation, Eq.~\eqref{eq:selfenergy}.
We finally get, for the functions $\gamma(\omega)$ and $\eta(\omega)$, 
\begin{align}
    \label{eq:gammaNLO}
        \gamma(\omega) &= M^2 - \frac{2M_0^4}N \frac{\tilde\lambda^2(3+2\tilde\lambda)}{1+\tilde\lambda} \frac{M_0^2 (3+2\tilde{\lambda})}{\omega^2 + M_0^4 (3+2\tilde{\lambda})^2} \\
    \label{eq:etaNLO}
        \eta(\omega) &= 1 + \frac{2M_0^4}N \frac{\tilde\lambda^2(3+2\tilde\lambda)}{1+\tilde\lambda} \frac1{\omega^2 + M_0^4 (3+2\tilde{\lambda})^2}
\end{align}
 As in the previous case, the response function and the field correlator can be decomposed as a sum of two poles, see Eq.~\eqref{eq:propagFouriertwoloop}. At the present order of approximation, we get
\begin{align}
       \Lambda_{1,1} &= M^2 \qty[ 1 - \frac{1}{N} \frac{\tilde\lambda^2(3+2\tilde\lambda)}{(1+\tilde{\lambda})^2} + \order{\frac1{N^2}}] \\
       \Lambda_{3,1}  &= M^2 \qty[ 3 + 2\tilde{\lambda}  + \order{\frac1{N}}]
    \label{}
\end{align}
and
\begin{align}
    c^\varphi_1&=1-\frac{\tilde\lambda^2(3+2\tilde\lambda)}{2N(1+\tilde{\lambda})^3}+ \order{\frac1{N^2}}\\
    c^\varphi_3 &= \frac{\tilde\lambda^2(3+2\tilde\lambda)}{2N(1+\tilde{\lambda})^3} + \order{\frac1{N^2}}
    \label{eq:cpmN}
\end{align}
Similarly to the previous perturbative calculation, the coefficient $c^\varphi_3$ being of order $1/N$, we only obtain the LO expression for $\Lambda_{3,1}$. 

Let us now consider the $\ev{\chi\chi}$ correlator which, at LO, is simply given by the infinite chain of bubbles. Indeed, one easily shows (see Appendix \ref{sec:largeN2}) that 
\begin{equation}
    C_{12}(\omega) = - \frac{3}{\lambda N}\mathbb{I}_{12}(\omega).
    \label{eq:phi2phi2I}
\end{equation}
From this, we get the (connected) correlator of the composite field  $\chi = \varphi^2/(2N)$
\begin{equation}
    G_{\chi\chi}(\omega)= \frac{1}{2 N M_0^2 }G_{2M_0^2(1+\tilde\lambda)}(\omega)+\order{\frac1{N^2}}
    \label{eq:rhorho}
\end{equation}
and we deduce the LO expressions
\begin{align}
  \Lambda_{2,0} &= 2 M_0^2(1+\tilde{\lambda}),\\
 c^\chi_2&=\frac{2}N(1+\tilde\lambda).
\end{align}

We finally need to solve Eq.~\eqref{eq:gap} for the local contribution $M^2$ at NLO. To this aim, we use 
\begin{equation}
\gamma(0)=M^2\left[1 - \frac{2}N \frac{\tilde\lambda^2}{ 1+\tilde{\lambda}}+\order{\frac1{N^2}}\right],
\end{equation}
from which we obtain
\begin{equation}
    M^2 = M_0^2 \qty[1+\frac2N \frac{\tilde{\lambda} (1 + \tilde{\lambda} + \tilde{\lambda}^2)}{(1+\tilde{\lambda})^2} + \order{\frac1{N^2}}].
    \label{eq:selfconsistentmassN}
\end{equation}
Collecting the previous results, we have, for the dynamical mass,
\begin{equation}
m_{\rm dyn}^2=M_0^2\left[1 + \frac{2}N \frac{\tilde{\lambda}}{ (1+\tilde{\lambda})^2}+ \order{\frac1{N^2}}\right]
\end{equation}
and for the lowest eigenvalues
\begin{align}
    \label{eq:lambda1NLO}
        \Lambda_{1,1}  &= M_0^2 \qty[ 1 +  \frac{1}{N}\frac{\tilde\lambda(2-\tilde\lambda)}{(1+\tilde{\lambda})^2}  + \order{\frac1{N^2}}], \\
    \label{eq:lambda2NLO}
        \Lambda_{2,0}  &= M_0^2\qty[2+2\tilde{\lambda} + \order{\frac1N}], \\
    \label{eq:lambda3NLO}
        \Lambda_{3,1}  &= M_0^2 \qty[ 3 + 2\tilde{\lambda}  + \order{\frac1{N}}].
\end{align}

As for the previous perturbative expressions, the above results exactly agree with those of the direct QFT calculations in Ref.~\cite{Gautier:2015pca}. In fact the agreement concerns all the intermediate quantities $\Pi$, $\mathbb I$ and $\Sigma$, using the rescalings \eqref{eq:rescaling} of the parameters and 
\begin{equation}
    \hat G = \frac{d\Omega_{D+1}}2 G \qc \hat{\mathbb{I}} = \frac{\Omega_{D+1}}2 \mathbb{I}, \qand \hat \Sigma = \frac{\Omega_{D+1}}{2d} \Sigma
    \label{}
\end{equation}
for the different two-point functions. The very same results have also been recently obtained from a QFT calculation in Euclidean de Sitter in Ref.~\cite{LopezNacir:2019ord}. That such very different calculations agree is a nontrivial result. Such an agreement between the stochastic approach and direct QFT calculations on either Lorentzian or Euclidean de Sitter was already well-known for equal-time correlators, {\it e.g.}, $\ev{\varphi^n}$, which measure the local field fluctuations \cite{Tsamis:2005hd,Rajaraman:2010xd,Beneke:2012kn}. Although expected on the basis of general arguments \cite{Tsamis:2005hd,Garbrecht:2013coa,Garbrecht:2014dca}, the agreement mentioned here for unequal time (nonlocal) correlators is far less trivial, in particular, for nonperturbative approximation schemes, and the present results, together with those of Refs.~\cite{Gautier:2015pca} and \cite{LopezNacir:2019ord} provide an explicit nontrivial check.

\subsection{Discussion}

\begin{figure}[t]
    \centering
    \includegraphics[width=\linewidth]{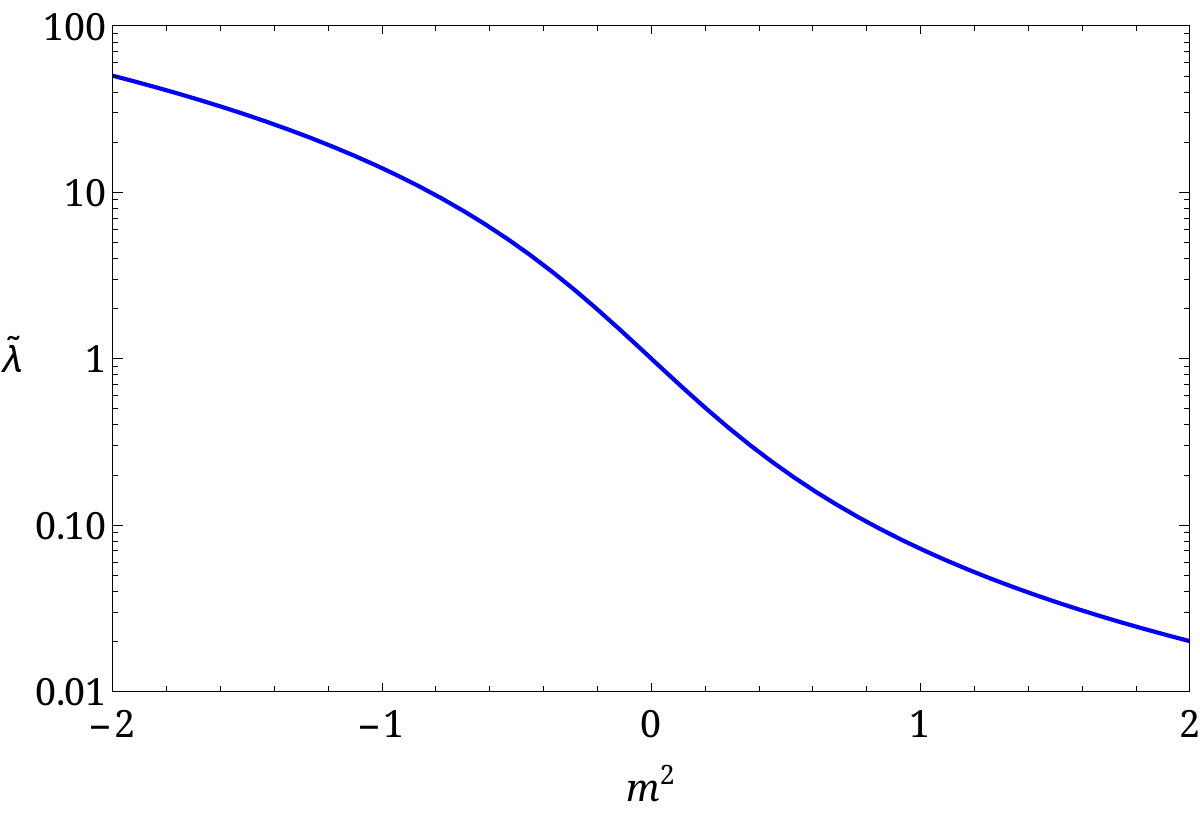}
    \caption{Effective coupling $\tilde \lambda$ as a function of the bare squared mass $m^2$. The bare coupling is taken as $\lambda=1$. The coupling becomes strongly nonperturbative for small and negative values of $m^2$.}
    \label{fig:lambda}
\end{figure}

We now discuss the results we obtained for the eigenvalues and associated correlators in several regimes. First, we check that the expressions we have for the $\Lambda_{n,\ell}$ and $c^{\varphi,\chi}_n$ coincide in the limit where we take both $N$ large and a the coupling $\bar \lambda$ small. In this regime, introducing $\bar\lambda_\infty=\lim_{N\to\infty}\bar\lambda=\lambda/(12m^4)$, we have $M_0^2=m^2[1+\bar\lambda_\infty-\bar\lambda_\infty^2+\order{\bar\lambda_\infty^3}]$ and $\tilde\lambda=\bar\lambda_\infty+\order{\bar\lambda_\infty^2}$, thus the two effective coupling coincide. For example, it is easy to check that Eq.~\eqref{eq:lambda1threeloop} coincides with the first member of Eq.~\eqref{eq:lambda1NLO} to give 
\begin{align}
        \frac{\Lambda_{1,1}}{m^2}&= 1 + \bar \lambda_\infty - \bar \lambda_\infty^2 + 2\bar \lambda_\infty^3+ \frac{2 \bar \lambda_\infty - 7 \bar \lambda_\infty^2 + 27 \bar \lambda_\infty^3}{N} \nonumber\\
        &+ \order{\bar\lambda_\infty^4,\frac1{N^2}}.
    \label{}
\end{align}

\begin{figure}[t]
    \centering
    \includegraphics[width=\linewidth]{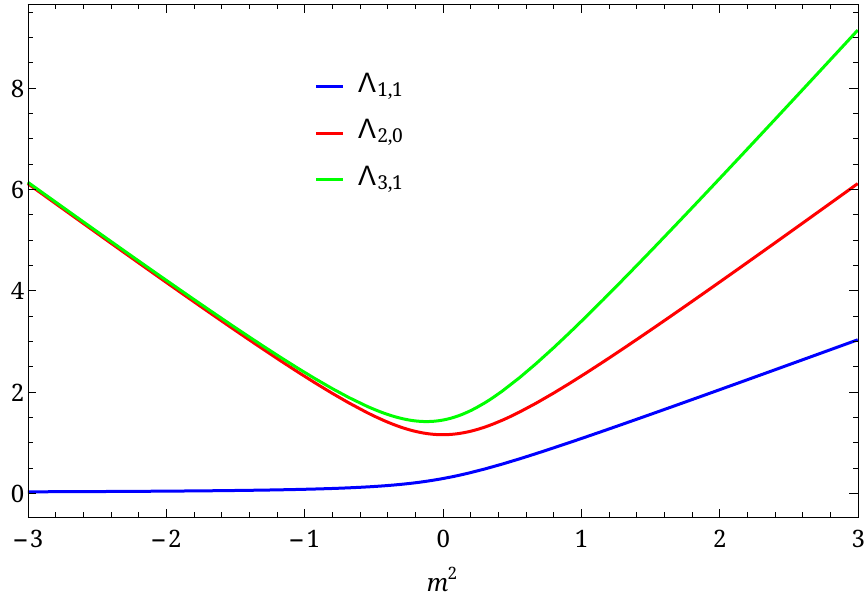}
    \caption{The eigenvalues $\Lambda_{1,1}$, $\Lambda_{2,0}$ and $\Lambda_{3,1}$ at leading order in the $1/N$ expansion as functions of the tree-level square mass $m^2$, for $\lambda=1$.}
    \label{fig:eigenvalues}
\end{figure}

The $1/N$ expansion allows the study of the nonperturbative regime in $\bar\lambda$, which correspond to either small or negative $m^2$ \cite{Gautier:2015pca}. This is illustrated in Fig.~\ref{fig:lambda}, where we show the effective coupling $\tilde\lambda$ as a function of $m^2$ for fixed coupling $\lambda$. The latter is of order one in the small mass regime $m^2=0$ and becomes large for $m^2<0$. Let us discuss the leading-order eigenvalues \eqref{eq:lambda1NLO}--\eqref{eq:lambda3NLO} in these regimes. The latter rewrite, in terms of the parameters $m^2$ and $\lambda$,
\begin{align}
        \Lambda_{1,1} &= \frac{m^2}2 + \sqrt{ \frac{m^4}4 + \frac\lambda{12} } , \\
        \Lambda_{2,0} &= 4 \sqrt{ \frac{m^4}4 + \frac\lambda{12} } , \\
        \Lambda_{3,1} &= \frac{m^2}2 + 5\sqrt{ \frac{m^4}4 + \frac\lambda{12} } .
    \label{}
\end{align}
and are plotted as functions of $m^2$ for $\lambda=1$ in Fig.~\ref{fig:eigenvalues}. 

In the small mass regime, we have
\begin{align}
       \Lambda_{1,1} = \sqrt{\lambda/12}\,,\quad\Lambda_{2,0} = 4\Lambda_{1,1}\,,\quad\Lambda_{3,1} = 5\Lambda_{1,1},
\end{align}
where we see that all eigenvalues are of the same order, so that all the correlators computed here have relatively large autocorrelation times, in particular, in the case of small coupling $\lambda\ll1$. This reflects the fact that the potential is very flat in that case.

Instead, in the case of a steep symmetry breaking tree-level potential, with $m^2<0$ and $\lambda/m^4\ll1$, we have
\begin{align}
\label{eq:lambda1m2neg}
       \Lambda_{1,1} &= \lambda/(12|m^2|)  \\
\label{eq:lambda23m2neg}
       \Lambda_{2,0} &= \Lambda_{3,1}=2 |m^2|\gg\Lambda_{1,1} .
\end{align}
The presence of a small ($\Lambda_{1,1}$) and a large ($\Lambda_{3,1}$) eigenvalue in the correlator of the vector field $\varphi$ reflects the existence of a flat (Goldstone mode) and a steep (Higgs mode) direction in the tree-level potential.\footnote{The absence of a true Goldstone mode in the actual spectrum, $\Lambda_{1,1}\neq0$, is due to the effective symmetry restoration by the infrared modes \cite{Ford:1985qh,Ratra:1984yq,Serreau:2011fu,Serreau:2013eoa}.} The eigenvalue $\Lambda_{2,0}$ is, again the longitudinal Higgs mode, the only one which contributes to the correlator of the  field $\chi$.

Interestingly, the present large-$N$ results share similarities with similar analytical results in the case $N=1$ in the limit of a steep double-well potential \cite{Starobinsky:1994bd,Markkanen:2020bfc}. Intuitively, when the two minima are far apart, the situation can be described as a superposition of two single-well spectra with tunnel effect yielding  infinitesimally split energy levels. Because the ground (equilibrium) state has $\Lambda_0=0$, this results in an exponentially suppressed, instanton-like value of $\Lambda_1\propto\exp(-a/\lambda)$, with $a$ a positive constant. Higher eigenvalues are essentially those of the unperturbed separate Gaussian wells with curvature $2|m^2|$, which yields pairs of quasidegenerate levels with eigenvalues $2 n |m^2|$, with $n\in\mathds{N}$ \cite{Starobinsky:1994bd}. Moreover, as pointed out in Ref.~\cite{Markkanen:2020bfc} the actual effective potential which enters the relevant Fokker-Planck equation is actually a three-well potential and there exist, consequently, additional states with eigenvalues $(n+1)|m^2|$, corresponding to the central well around $\phi=0$. There are two major differences in the case of a continuous symmetry $N>1$. First, the presence of flat directions (Goldstone modes) in the potential, results in a milder, power law suppression for the first nonzero eigenvalue $\Lambda_{1,1}$, see Eq.~\eqref{eq:lambda1m2neg}. Interestingly, we observe that the $1/N$ expansion becomes singular for the latter when $N\to1$. In the limit of a steep symmetry breaking potential $\tilde\lambda\gg1$, we have  
\begin{equation}
    \Lambda_{1,1} = \frac\lambda{12\abs{m^2}}\qty[1-\frac1N + \order{\frac1{N^2}}],
    \label{}
\end{equation}
and similarly for the coefficient $c^\varphi_1=1-1/N + \order{1/N^2}$. The second difference with the case $N=1$ is that, for sufficiently large $N$, the central well is lifted by a factor $\propto N$ and the corresponding excitations decouple \cite{Moreau2020}. There remains only the quasidegenerate levels with eigenvalues $2 n |m^2|$, corresponding to the heavy longitudinal directions in the potential, see Eq.~\eqref{eq:lambda23m2neg}.

In terms of correlation functions, this implies that in the case $m^2<0$ the correlation time in the  vector channel ($\varphi$) is considerably larger than in the scalar channel ($\chi$), which does not see the flat transverse direction. It is to be expected that such large correlation times also occur for composite fields in higher representations (tensor channels). These correlation times are related to other quantities of physical interest, such as the relaxation (or equilibration) times from an excited state to the BD vacuum, decoherence timescales \cite{Giraud:2009tn,Gautier:2012vh}, or, closest to standard phenomenological interest, to the spectral index of various observables \cite{Starobinsky:1994bd,Markkanen:2019kpv}. Exploiting de Sitter invariance, the spectral index of a given field ${\cal A}$ can be read off the decomposition \eqref{eq:AAdecomp} as $n_{\cal A}-1 = 2 \Lambda_{\cal A}$, with $\Lambda_{\cal A}$ the lowest eigenvalue contributing to the sum \eqref{eq:AAdecomp}. For instance, the spectral index of the field $\varphi$ is given by $n_\varphi-1=2\Lambda_{1,1}$. Similarly, $\Lambda_{2,0}$ is related to the spectral index of the field $\chi$ or of other typical O($N$) scalar quantity. For instance, as discussed in Ref.~\cite{Markkanen:2019kpv}, the density contrast $\delta= (V - \ev{V})/\!\ev{V}$ has a spectral index $n_\delta-1=2\Lambda_{2,0}$.

Finally, we mention an interesting role played by the eigenvalue $\Lambda_{3,1}$ based on the discussion in Sec.~\ref{sec:noisecorrelator}. According to Eqs.~\eqref{eq:etatwoloop} and \eqref{eq:etaNLO}, we have
\begin{equation}
        \eta(\omega) = 1 + 2c \frac{\Lambda_{1,1}\Lambda_{3,1}}{\omega^2 +  \Lambda_{3,1}^2}
\end{equation}
both in the perturbative and in the $1/N$ expansions, with 
\begin{equation}
 c=\frac{\bar\lambda^2}{N+2} +\order{\bar\lambda^3}=\frac{1}N \frac{\tilde\lambda^2}{1+\tilde\lambda} +{\cal O}\qty(N^{-2}).
\end{equation}
In real time, this gives
\begin{equation}
    \eta(t) = \delta(t) + c\Lambda_{1,1}e^{-\Lambda_{3,1}|t|}.
\end{equation}
With Eq.~\eqref{eq:noiseandsigma}, we see that $\Lambda_{3,1}^{-1}$ is the autocorrelation time of the effective colored noise correlation in the vector channel due to the infrared modes while $c\Lambda_{1,1}$ controls the amplitude of the colored contribution.  In the perturbative regime $m^2>0$, the autocorrelation time is small $\sim 1/m^2$ with small amplitude $\sim\lambda^2/m^2$. However the autocorrelation time can be either parametrically large $\sim1/\sqrt\lambda$ with a small amplitude $\sim\sqrt\lambda$ for $m^2=0$, or small $\sim 1/|m^2|$ with ``large'' amplitude $\sim1$ for $m^2<0$.

\begin{figure}[t]
    \centering
    \includegraphics[width=\linewidth]{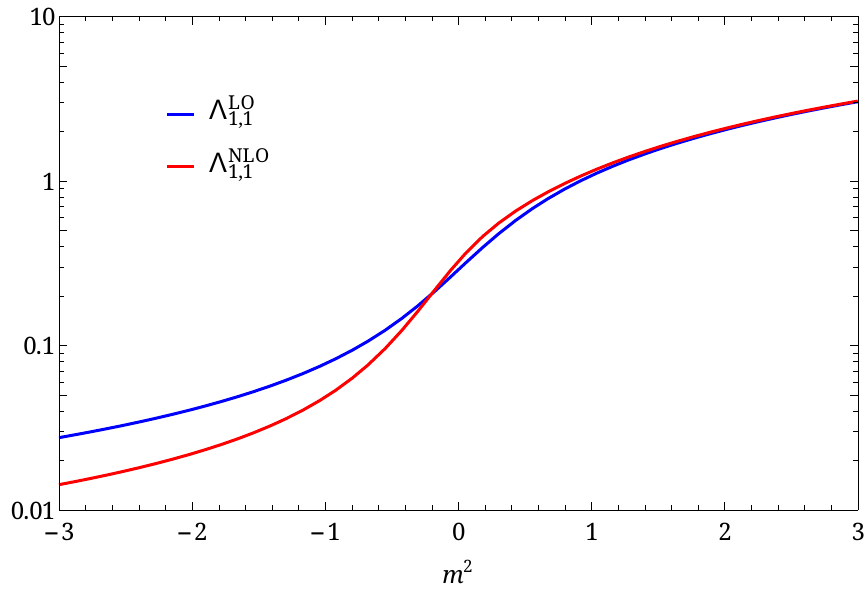}
\caption{The lowest nonzero eigenvalue $\Lambda_{1,1}$ at leading and next-to-leading orders in the $1/N$ expansion as a function of $m^2$ with $\lambda=1$, and for $N=2$. }
    \label{fig:LOvsNLO}
\end{figure}

We close this section by comparing the expressions of $\Lambda_{1,1}$ at leading and next-to-leading orders in the $1/N$ expansion as a function of the parameters of the theory for the extreme case $N=2$ in Fig.~\ref{fig:LOvsNLO}.

\section{Conclusions}

\label{sec:Concl}

To conclude, we used the JdD path integral formulation of the stochastic equation that describes the infrared dynamics of an $O(N)$ theory of test scalar fields to study the two-point unequal time correlators of various operators. The resulting field theory is a one-dimensional supersymmetric theory with $N$ scalar superfields. This supersymmetry is a mere consequence of the symmetries of the original system in stationary state, and was used to show that the correlators of the various fields are not independent. One of the obtained relations can be interpreted as a fluctuation-dissipation relation, showing the analogy of the system with a Brownian motion with a thermal noise at de Sitter temperature \cite{Rigopoulos:2016oko}.

Having in mind the result of the Fokker-Planck formulation, which is usually solved as an eigenvalue problem \cite{Starobinsky:1994bd,Markkanen:2019kpv}, we then discussed the general structure of the unequal time two-point correlator of composite operators of the scalar field. It can be expressed as a sum of free propagator with a hierarchy of mass scales, which corresponds to a subset of the tower of eigenvalues previously mentioned. We have computed explicitly the $\ev{\varphi\varphi}$ and $\ev{\chi\chi}$ correlators in two specific limits, the perturbative case up to three loop and the $1/N$ expansion at NLO, and we have obtained the values of the first three eigenvalues in both cases. We  have checked that our results coincide with other computations, done either in the Lorentzian \cite{Gautier:2013aoa,Gautier:2015pca} or Euclidean \cite{LopezNacir:2019ord} field theory. 

The result from the $1/N$ expansion is particularly interesting as it allows us to probe nonperturbative regimes. Such regimes corresponds to the massless and symmetry breaking case, the latter being difficult to probe numerically. In the limit of a deeply broken initial potential, we find that the lowest nonzero eigenvalue is strongly suppressed with respect to higher order ones. This has direct physical consequence, {\it e.g.}, in terms of equilibration times or power spectra of fields in the different representations of the O($N$) group. For instance, the vector channel $\ell=1$ has typically long range spacetime correlations whereas the scalar channel $\ell=0$ is typically (sometimes significantly) of shorter range.

There are several directions to extend the present analysis. First, we used the computation of correlators to access the mass scale hierarchy. When combined with our expansion schemes, this only gives the first eigenvalues due to the coefficients appearing in the sums of free propagators of the correlators. Alternative formulations, directly at the level of the Fokker-Planck equation, may be able to grasp the full hierarchy directly. 

On a more speculative level, this work is limited to test scalar fields, and an important question would be to extend the present considerations to a more realistic inflationary setup. An interesting intermediary step would be to consider systems with derivative interactions, {\it e.g.} along the lines of  Ref.~\cite{Kitamoto:2018dek}.

\section*{Acknowledgements}

We are grateful to T. Prokopec, G. Rigopoulos, V. Vennin, A. Rajantie and T. Markkanen for useful discussions at the various stages of this work.

\appendix

\section{Quantum vs stochastic field theory}
\label{sec:quantumvsstochastic}

Here, we discuss the relation between the various two-point functions of the stochastic theory and the underlying QFT. In particular, the spectral function of the QFT encodes information about the equal-time commutation relation which is lost in the slow-roll approximation yielding the effective Langevin equation \eqref{eq:stochastic1}. We show how this constraint is deleted in the process of coarse graining over the relevant timescales of the slow-roll regime and is replaced by the constraint \eqref{eq:rhoone} for the stochastic spectral function. To this aim we consider the stochastic theory (for a single free field, $N=1$, in the light mass limit) before taking the slow-roll limit. 

The effective stochastic theory applies to a  field $\hat\varphi(t,\vec x)$ spatially smeared on the scale of a Hubble patch. As recalled in Sec.~\ref{sec:setup}, integrating out the short wavelength, subhorizon mode leads to an effective Langevin equation for the infrared (smeared) field with the following essential features: First, the spatial gradients of the smeared field can be neglected, which results in effectively independent Hubble patches with only one (quantum mechanical) degree of freedom $\hat\varphi(t)$; Second, the quantum fluctuations of the latter can be described by those of a classical stochastic field sourced by a random noise which reflects the effect of the ultraviolet modes; Finally, in the BD vacuum, the latter is a Gaussian white noise. A simple model Langevin equation is 
\begin{equation}\label{appeq:langevinmodel}
 \ddot{\hat\varphi}+d\dot{\hat\varphi}+\hat m^2\hat\varphi=d\hat \xi,
\end{equation}
where the Gaussian noise correlator is [see Eq.~\eqref{eq:noisestochastic1}]
\begin{equation}\label{appeq:noisecor}
 \ev{\hat\xi(t)\hat\xi(t')}=\frac{2}{d\Omega_{D+1}}\delta(t-t').
\end{equation}
Obviously, the slow-roll limit \eqref{eq:stochastic1} is obtained by neglecting the term $\ddot{\hat\varphi}$. We shall make this more precise below.

\subsection{Retarded and spectral functions}

The spectral function of the quantum mechanical degree of freedom $\hat\varphi$ is defined as
\begin{equation}
 \hat\rho(t-t')=i\ev{\qty[\hat\varphi(t),\hat\varphi(t')]}
\end{equation}
and is normalized through the equal-time commutation relation
\begin{equation}\label{appeq:normrho}
 \dot{\hat\rho}(t=0)=i\ev{\qty[\dot{\hat\varphi}(t),\hat\varphi(t)]}=Z.
\end{equation}
Here, we allow for an arbitrary normalization of the smeared field. 
The retarded Green function is defined as 
\begin{equation}
 \hat G_R(t,t')=\theta(t-t')\hat \rho(t-t')
\end{equation}
and solves the following equation
\begin{equation}
 \qty(\partial_t^2+d\partial_t+m^2)\hat G_R(t)=Z\delta(t).
\end{equation}

In frequency space, we have
\begin{equation}
 \hat G_R(\omega)=\frac{Z}{-\omega^2-id\omega+\hat m^2}
\end{equation}
and, for the spectral function, 
\begin{equation}
 \hat \rho(\omega)=2i\Im \hat G_R(\omega)=\frac{2idZ\omega}{\qty(\omega^2-\hat m^2)^2+(d\omega)^2}.
\end{equation}
In real time, this gives the standard expression for the (over)damped harmonic oscillator
\begin{equation}
 \hat\rho(t)=Z\frac{\sinh(\nu t)}{\nu}e^{-\frac d 2 |t|},
\end{equation}
with $\nu=\sqrt{d^2/4-\hat m^2}$. It is useful to rewrite these expressions in terms of the roots of the inverse retarded function $G_R^{-1}(-i\omega_\pm)=0$, that is, $\omega_\pm=d/2 \pm \nu$. We have
\begin{equation}\label{appeq:grfreq}
 \hat G_R(\omega)=\frac{Z}{2\nu}\qty(\frac{i}{\omega+i\omega_-}-\frac{i}{\omega+i\omega_+}),
\end{equation}
\begin{equation}\label{appeq:rhofreq}
 \hat \rho(\omega)=\frac{Z}{2\nu}\qty(\frac{2i\omega}{\omega^2+\omega_-^2}-\frac{2i\omega}{\omega^2+\omega_+^2}),
\end{equation}
and
\begin{align}
\hat \rho(t)=\frac{Z}{2\nu}{\rm sign}(t)\qty(e^{-\omega_-|t|}-e^{-\omega_+|t|}).
\end{align}

\subsection{Statistical correlator}

The statistical correlator of the quantum field is defined as 
\begin{equation}\label{appeq:statcor}
 \hat F(t,t')=\frac{1}{2}\ev{\hat\varphi(t)\hat\varphi(t')+\hat\varphi(t')\hat\varphi(t)}.
\end{equation}
The effective stochastic theory relies on the assumption that the smeared infrared field is essentially classical. At the level of the field correlators, this means that the spectral function (which encodes the noncommuting aspects of the quantum fields) is small compared to the statistical correlator (which encodes the typical occupation number) \cite{Aarts:2001yn}, that is, $\hat \rho\ll\hat F$. In that case, the statistical correlator \eqref{appeq:statcor} reduces to the stochastic correlator 
\begin{equation}
 \hat F(t,t')\approx G_{\hat\varphi\hat\varphi}(t,t'),
\end{equation}

In the present simple model, the latter is obtained as follows. Equation \eqref{appeq:langevinmodel} is solved as  
\begin{equation}
 \hat\varphi(t)= \hat \varphi_0(t)+\frac{d}{Z}\int dt'\hat G_R(t-t')\xi(t'),
\end{equation}
where the solution of the homogeneous equation reads
\begin{equation}
  \hat \varphi_0(t)=A_-e^{-\omega_-t}+A_+e^{-\omega_+t},
\end{equation}
with $A_\pm$ arbitrary constants of integration. The latter describes the transient regime from generic initial conditions to the late-time equilibrium. The equilibrium stochastic correlator is obtained as, in frequency space, 
\begin{equation}
 G_{\hat\varphi\hat\varphi}(\omega)=\frac{d^2}{Z^2}{\cal N}(\omega)|\hat G_R(\omega)|^2,
\end{equation}
where $\hat{\cal N}(\omega)=2/(d\Omega_{D+1})$ is the Fourier transform of the noise correlator \eqref{appeq:noisecor}. We get 
\begin{equation}\label{appeq:gfffreq}
 G_{\hat\varphi\hat\varphi}(\omega)=\frac{1}{\Omega_{D+1}\nu}\qty(\frac{1}{\omega^2+\omega_-^2}-\frac{1}{\omega^2+\omega_+^2}),
\end{equation}
or, equivalently, in real time,
\begin{equation}\label{appeq:gfftime}
 G_{\hat\varphi\hat\varphi}(t)=\frac{1}{\Omega_{D+1}\nu}\qty(\frac{e^{-\omega_-|t|}}{2\omega_-}-\frac{e^{-\omega_+|t|}}{2\omega_+}).
\end{equation}

\subsection{Fluctuation-dissipation relation}

The spectral function and the statistical correlator satisfy the identity 
\begin{align}\label{appeq:fd}
 \hat\rho(\omega)=i\omega Z\Omega_{D+1} G_{\hat\varphi\hat\varphi}(\omega),
\end{align}
or, in real time,
\begin{align}
 \hat\rho(t)=-Z\Omega_{D+1} \partial_t G_{\hat\varphi\hat\varphi}(t),
\end{align}
which is characteristic of a thermally equilibrated system in the high temperature (low frequency) limit. Indeed, in a thermal state with inverse temperature $\beta$, one has the following exact relation between the statistical correlator (fluctuation) and the spectral function (dissipation)
\begin{equation}\label{appeq:fd2}
 G_{\hat\varphi\hat\varphi}(\omega)=\frac{\hat\rho(\omega)}{2i\tanh (\beta\omega/2)}\to\frac{\hat\rho(\omega)}{i\beta\omega},
\end{equation}
where the last expression is obtained in the limit ${\beta\omega\ll1}$. This also corresponds to the classical field regime, where the occupation number $n(\omega)=[\exp(\beta\omega)-1]^{-1}\to1/(\beta\omega)\gg1$. Comparing Eqs.~\eqref{appeq:fd} and \eqref{appeq:fd2}, we read the inverse temperature\footnote{Note, that with the choice $Z^{-1}={\cal V}_d=\Omega_d/d$, the volume of a spherical Hubble patch (of radius $H^{-1}=1$), we get $ \beta=2\pi$, the Gibbons-Hawking temperature of de Sitter space. } $\beta=Z\Omega_{D+1}$.

\begin{figure}[t]
    \centering
    \includegraphics[width=\linewidth]{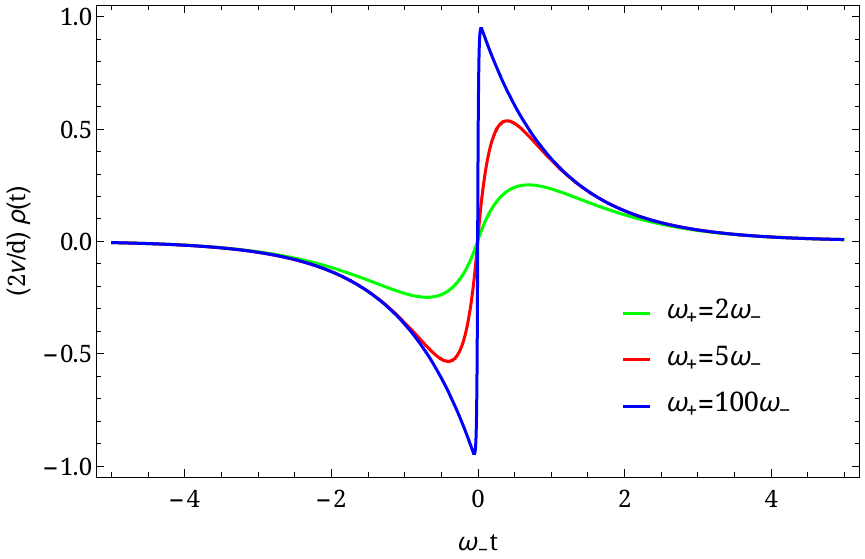}
    \includegraphics[width=\linewidth]{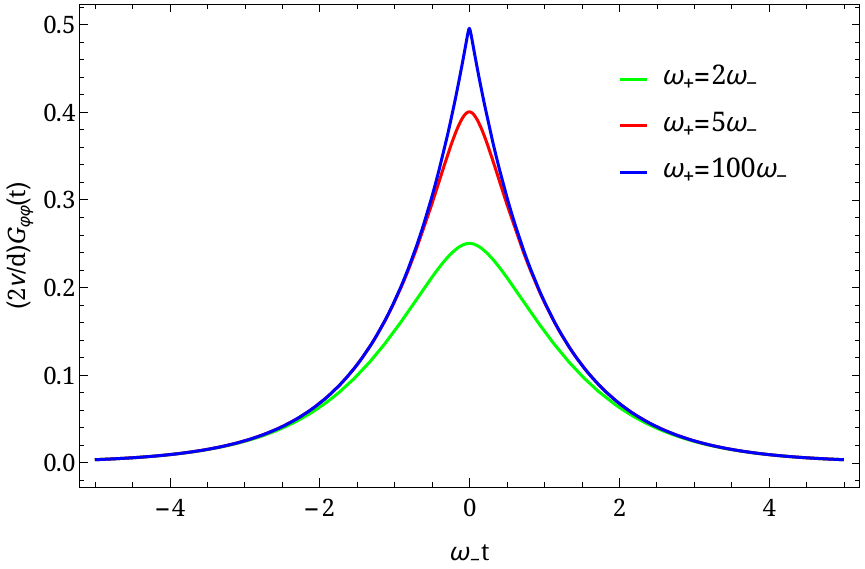}
    \caption{ The spectral function $\rho$ and the statistical correlator $G_{\varphi\varphi}$ (for the rescaled field $\varphi$) as functions of time in units of $\omega_-$ for different values of $\omega_+$. This illustrates the coarse graining of the short-time structure as a clear separation of scales $\omega_-\ll\omega_+$ arises. We have rescaled both functions by a factor $2\nu/d=(\omega_+-\omega_-)/(\omega_++\omega_-)\to1$ for a nicer plot.} 
    \label{fig:slowroll}
\end{figure}

\subsection{Slow-roll limit}

We are now in a position to clearly identify the necessary requirement for the slow-roll limit, which, we recall, amounts to neglecting the term $\ddot{\hat\varphi}$ in Eq.~\eqref{appeq:langevinmodel}. We see that, in that case, the only root of the homogeneous equation is $\omega=-i\hat m^2/d$. This corresponds to, first, taking the small mass limit $\hat m^2\ll1$, so that $\omega_-\approx \hat m^2/d\ll\omega_+\approx d$ and, second, keeping only the contribution from the lowest pole in Eqs.~\eqref{appeq:grfreq}, \eqref{appeq:rhofreq}, and \eqref{appeq:gfffreq}. In real time, this amounts to neglecting $e^{-\omega_+|t|}\ll e^{-\omega_-|t|}$, which amounts to a coarse graining on timescales $\sim\omega_-^{-1}$.

We define the stochastic response and spectral functions\footnote{From their definitions, these are insensitive to a change of normalization so that, {\it e.g.}, $G_{\varphi\tilde\varphi}=G_{\hat\varphi\tilde{\hat\varphi}}$. In other words, the response field $\tilde\varphi$ rescales as the inverse of $\varphi$.} $G_{\varphi\tilde\varphi}$ and $\rho$ as
\begin{equation}
 \hat G_R=  \frac{Z}{d}G_{\varphi\tilde\varphi}\quad{\rm and}\quad\hat\rho=\frac{Z}{d}\rho.
\end{equation}
In the slow-roll limit we thus have $G_{\varphi\tilde\varphi}(t)=\theta(t)\rho(t)$, with
\begin{equation}\label{appeq:rho}
 \rho(t)\to{\rm sign}(t)e^{-\frac{\hat m^2}{d}|t|}.
\end{equation}

It is important to notice that in this coarse-graining process, we lose the property \eqref{appeq:normrho}, for which the presence of both roots $\omega_\pm$ is essential. In  other words, the coarse-grained, stochastic spectral function does not resolve the short-time structure close to $t=0$. However, the constraint  \eqref{appeq:normrho} from quantum mechanics is replaced by another one in the coarse-grained theory. Indeed, one easily checks in the present simple example that for $\hat m^2\to0$, the maximal value of the QFT spectral function is $\hat\rho_{\rm max}\to Z/d$, that is, $\rho_{\rm max}\to 1$, which occurs at $t_{\rm max}\to d^{-1}\ln (d^2/\hat m^2)$. In units of the stochastic timescale $\omega_-^{-1}=d/\hat m^2$, we have $\omega_-t_{\rm max}= (\hat m^2/d
^2)\ln (d^2/\hat m^2)\to0^+$. We thus have 
\begin{equation}
\label{appeq:stochrho1}
 \rho(t=0^+)=1,
\end{equation}
where $t\to0^+$ is to be understood in units of the relevant timescale $\omega_-^{-1}$. The identity \eqref{appeq:stochrho1} also follows directly from the expression \eqref{appeq:rho}.

Finally, the stochastic correlator \eqref{appeq:gfftime} becomes, in the slow-roll limit, 
\begin{equation}
 G_{\hat\varphi\hat\varphi}(t)\to\frac{e^{-\frac{\hat m^2}{d}|t|}}{\Omega_{D+1}\hat m^2},
\end{equation}
which corresponds to Eq.~\eqref{eq:freeprop} after the proper rescaling of the field $G_{\hat\varphi\hat\varphi}(t)=\frac{2}{d\Omega_{D+1}}G_{\varphi\varphi}(t)$. Here, we see that in the slow-roll limit, we automatically satisfy the classicality condition $G_{\hat\varphi\hat\varphi}\gg\hat\rho$ mentioned above. We illustrate the coarse-graining process in Fig.~\ref{fig:slowroll} by showing the stochastic spectral function and correlator for various values of $\hat m^2$ in units of the stochastic timescale $\omega_-$. We clearly see how the ultraviolet, short-time structure near $t=0$ is washed out, leading to a singular behavior. We also see the condition \eqref{appeq:stochrho1} emerging.

We end by remarking that the fluctuation-dissipation relation discussed above remains valid in the slow-roll limit and reads
\begin{align}
 \rho(t)=-d\Omega_{D+1} \partial_t G_{\hat\varphi\hat\varphi}(t),
\end{align}
which is nothing but the relation \eqref{eq:fdrealtime} for the particular model considered here, again with the appropriate rescaling of the fields.

\section{Two-loop order}
\label{sec:perturbative3}

\begin{figure}[t]
    \centering
    \includegraphics{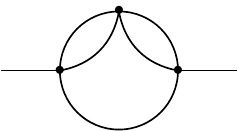}
    \caption{Three-loop contribution to the self-energy $\Sigma$.} 
    \label{fig:perturbative_appendix}
\end{figure}

Here, we detail the computation of the superfield self-energy up to two-loop order in the perturbative expansion. Standard diagrammatic rules yield, for the one-loop contribution, given by the diagram (b) of Fig.~\ref{fig:perturbative}
\begin{equation}
    \Sigma^{(2a)}_{12} =  \delta_{12}\frac{N+2}{3N} \frac{i\lambda}{2} \int \frac{\dd\omega}{2\pi} G^{m^2}_{12}(\omega).
    \label{}
\end{equation} 
Using the expression \eqref{eq:treeprop} of the tree-level propagator and the relations 
\begin{align}
\label{eq:relations1}
        \delta_{12}^2 &= 0 ,\\
        \delta_{12} \qty(K_\omega\delta_{12}) &= \delta_{12} ,\\
\label{eq:relations3}
        \qty(K_\omega\delta_{12})\qty( K_{\omega'}\delta_{12}) &= K_{\omega+\omega'}\delta_{12},
\end{align}
we easily perform the Grassmann algebra to get
\begin{equation}
    \Sigma^{(2a)}_{12} = i\delta_{12} \frac{N+2}{3N} \frac{\lambda}{2} {\cal F}(m^2),
    \label{}
\end{equation}
where we defined
\begin{equation}
{\cal F}(m^2)= \int \frac{\dd\omega}{2\pi} \frac1{\omega^2+m^4}=\frac{1}{2m^2}.
\end{equation}
This yields Eq.~\eqref{eq:sigma1looppert}. The local two-loop contribution given by the diagram (b) of Fig.~\ref{fig:perturbative} reads, writing $\int_3=\int \dd{\theta_3}\dd{\bar\theta_3}$,
\begin{align}
        \Sigma^{(2b)}_{12} &= \frac{\lambda^2}{4} \qty(\frac{N+2}{3N})^2 \int_{3} \int \frac{\dd\omega}{2\pi} G^{m^2}_{13}(\omega) G^{m^2}_{32}(\omega) \\
        & \qquad \times \int \frac{\dd\omega'}{2\pi} G^{m^2}_{33}(\omega') \delta_{12}.
    \label{}
\end{align}
Using Eqs.~\eqref{eq:relations1}-\eqref{eq:relations3} again, we get
\begin{equation}
    \Sigma^{(2b)}_{12} = i\delta_{12}\qty(\frac{N+2}{3N}\frac\lambda2)^2 {\cal F}'(m^2){\cal F}(m^2),
\end{equation}
which gives Eq.~\eqref{eq:sigma2looppert}. Finally, the nonlocal contribution at two loop is given by the diagram (c) of Fig.~\ref{fig:perturbative} and can be expressed as 
\begin{equation}
    \begin{aligned}
        \Sigma^{(2c)}_{12}(\omega) &= \frac{\lambda^2}6 \frac{N+2}{3N} \int \frac{\dd\omega'}{2\pi}\frac{\dd\omega''}{2\pi} G^{m^2}_{12}(\omega-\omega'-\omega'') \\
        & \qquad \times G^{m^2}_{12}(\omega') G^{m^2}_{12}(\omega'') .
    \end{aligned}
    \label{}
\end{equation}
We perform the frequency integrations using the following identity
\begin{equation}
    \int \frac{\dd\omega'}{2\pi} G^{m_A^2}_{12}(\omega') G^{m_B^2}_{12}(\omega-\omega') = \frac1{2m_{AB}^2} G^{m_A^2+m_B^2}_{12}(\omega) .
    \label{eq:relation2}
\end{equation}
where $m^2_{AB}$ is the reduced square mass, defined as
\begin{equation}
    \frac1{m_{AB}^2} = \frac1{m_A^2} + \frac1{m_B^2}.
    \label{appeq:nonloc}
\end{equation}
Equation \eqref{eq:relation2} expresses the fact that the product of two tree-level superpropagators in real time is proportional to a single superpropagator with the sum of the two square masses. For instance, for the component $G_{\varphi\varphi}$ of the propagator, see Eq.~\eqref{eq:freeprop}, we have, trivially,\footnote{A similar relation holds for tree-level propagators of quantum fields in Lorentzian \cite{Gautier:2013aoa,Gautier:2015pca} and Euclidean \cite{LopezNacir:2019ord} de Sitter in the appropriate limit.} 
\begin{equation}
 G_{m_A^2}(t)G_{m_B^2}(t)=\frac1{2m_{AB}^2}G_{m_A^2+m_B^2}(t).
\end{equation}
This generalizes to all the components of the superpropagator $G_{12}(t)$. Using this relation twice on the expression \eqref{appeq:nonloc} gives Eq.~\eqref{eq:selfenergy}.

\section{Three-loop order}
\label{sec:perturbative4}

We turn to the computation of $\Lambda_{3,1}$ at order $\order{\lambda}$. Indeed, we saw in Sec.~\ref{sec:perturbative} that taking diagrams up to order $\order{\lambda^2}$ only gives $\Lambda_{3,1}$ at LO because of the subleading coefficient $c^\varphi_3$. To circumvent this problem, we go to order $\order{\lambda^3}$, adding the diagram of Fig.~\ref{fig:perturbative_appendix}. Applying the techniques described in the previous section, we obtain, after some calculations, 
\begin{align}
       &\Sigma^{(10)}_{12}(\omega) =  -\frac{(N+2)(N+8)}{27 N^3} \frac{\lambda^3}{32 M^8} \nonumber\\
       &\times \left( 3 i M^2 \frac{\omega^2-27M^4}{(\omega^2+9M^4)^2}\delta_{12}  +\frac{\omega^2+45M^4}{(\omega^2+9M^4)^2}K_\omega\delta_{12} \right),
   \label{}
\end{align}
which gives the following expressions for $\gamma$ and $\eta$
\begin{align}
        \gamma(\omega) &= M^2\qty[1 -\frac{3 \alpha\lambda^2}{\omega^2+9M^4} \qty(1 + \beta\lambda\frac{\omega^2-27M^4}{\omega^2+9M^4}) ],\\
        \eta(\omega) &= 1 + \frac{\alpha\lambda^2}{\omega^2+9M^4} \qty(1 - \beta\lambda\frac{\omega^2+45M^4}{\omega^2+9M^4}) ,
    \label{}
\end{align}
where $\alpha=\frac{N+2}{3N^2} \frac1{8M^4}$ and $\beta=\frac{N+8}{9N}\frac1{4M^4}$. We then proceed as usual and compute the roots of $ -i G_{\varphi\tilde\varphi}^{-1}(\omega) = i\gamma(\omega)+\omega\eta(\omega) $, with
\begin{equation}
    -i G_{\varphi\tilde\varphi}^{-1}(\omega)  = \omega + iM^2 + \frac{\alpha\lambda^2}{\omega+3iM^2}\qty( 1 - \beta\lambda \frac{\omega+9iM^2}{\omega+3iM^2}) .
    \label{}
\end{equation}
To get the correct perturbative expression for the poles $\Lambda_{1,1}$ and $\Lambda_{3,1}$, and the coefficients $c_1$ and $c_3$, we have to factorize this expression such that the retarded propagator is decomposed into a sum of free propagators, as in Eq.~\eqref{eq:propagFouriertwoloop}. To do this, we write, up to higher-orders terms, 
\begin{equation}
    -i G_{\varphi\tilde\varphi}^{-1}(\omega) = \omega + iM^2 + \frac{\alpha\lambda^2(1-\beta \lambda)}{\omega+3iM^2(1+2\beta\lambda)} +\order{\bar\lambda^4}.
    \label{eq:Pomeg}
\end{equation}
This is the only combination compatible with the decomposition \eqref{eq:responsefraction} in simple fractions. Computing the poles and residue yields
\begin{align}
        \Lambda_{1,1} &= M^2 \qty( 1 - \frac\alpha{2 M^4} \lambda^2 + \frac{2\alpha\beta}{M^4} \lambda^3 ) ,\\
        \Lambda_{3,1} &= 3M^2 \qty( 1 + 2\beta\lambda) ,\\
        c^\varphi_1 &= 1 -  \frac{\alpha \lambda^2\qty(1 - \beta\lambda)}{4M^4}, \\
        c^\varphi_3 &= \frac{\alpha \lambda^2\qty(1 - 7\beta\lambda)}{4M^4}.
    \label{}
\end{align}

We now need to compute the effective square mass $M^2$ at three-loop order. This can be done either by a direct calculation of the relevant local contributions to the self-energy, or by following the strategy adopted in Sec.~\ref{sec:largeN}, that is, by solving the gap equation \eqref{eq:gap} at the appropriate order of approximation. We implement the latter here since this does not involve computing any new diagram. The exact gap equation for $M^2$ reads
\begin{equation}
 M^2=m^2+\frac{N+2}{3N}\frac{\lambda}{4\gamma(0)},
\end{equation}
where, at the present order of approximation,
\begin{align}
\gamma(0)=M^2\qty(1 - \frac{\alpha\lambda^2}{3M^4} +\frac{\alpha\beta\lambda^3}{M^4}).
\end{align}
This is readily solved as
\begin{equation}
\frac{M^2}{m^2}=1+\bar\lambda-\bar\lambda^2+2\frac{N+3}{N+2}\bar\lambda^3+\order{\bar\lambda^4}.
\end{equation}
We thus find, at three-loop order
\begin{align}
\frac{m_{\rm dyn}^2}{m^2}=1+\bar\lambda-\frac{N+4}{N+2}\bar\lambda^2+2\frac{N^2+9N+20}{(N+2)^2}\bar\lambda^3+\order{\bar\lambda^4}
\label{eq:mdynthreeloop}
\end{align}
and 
\begin{align}
 \frac{\Lambda_{1,1}}{m^2}&= 1 +\bar\lambda- \frac{N+5}{N+2}\bar\lambda^2 +\frac{2N^2+23N+62}{(N+2)^2}\bar\lambda^3+\order{\bar\lambda^4},
 \label{eq:lambda1threeloop}
\end{align}
together with 
\begin{align}
 c^\varphi_3=\frac{3\bar\lambda^2}{2(N+2)}-\frac{19N+80}{2(N+2)^2}\bar\lambda^3+\order{\bar\lambda^4}
\end{align}
and $c_1=1-c_3+\order{\bar\lambda^4}$.
We also get the ${\cal O}(\bar\lambda)$ correction to $\Lambda_{3,1}$
\begin{align}
 \frac{\Lambda_{3,1}}{3m^2}=1+\frac{5N+22}{3(N+2)}\bar\lambda+\order{\bar\lambda^2}.
 \label{lambda3threeloop}
\end{align}

\section{The $1/N$ expansion at NLO}
\label{sec:largeN2}

In this section, we give additional details on the computation of the self-energy and the $\ev{\chi\chi}$ correlator in the $1/N$ expansion. We begin with the self-energy. 

As previously mentioned, we can proceed in several steps. We first compute the one-bubble diagram $\Pi$ which is a mere convolution of two propagators. This is easily done with Eq.~\eqref{eq:relation2}. Putting this in the integral equation \eqref{eq:integeq}, we decompose each function in terms of the Grassmann structures 
\begin{align}
        \Pi_{12}(\omega) &= i\pi_\gamma(\omega) \delta_{12} +  \pi_\eta(\omega) K_{\omega}\delta_{12}, \\ 
    \label{appeq:defII}
        \mathbb I_{12}(\omega) &= iI_\gamma(\omega) \delta_{12} +  I_\eta(\omega) K_{\omega}\delta_{12},
\end{align}
and obtain the following equations:
\begin{align}
        I_\gamma(\omega) &= \pi_\gamma(\omega) -\pi_\gamma(\omega) I_\gamma(\omega) + \omega^2 \pi_\eta(\omega)I_\eta(\omega),\\
        I_\eta(\omega) &= \pi_\eta(\omega) - \pi_\gamma(\omega) I_\eta(\omega) - \pi_\eta(\omega) I_\gamma(\omega) .
\end{align}

Using 
\begin{align}
 \pi_\gamma(\omega)&=\frac{\lambda}{6M_0^2}\frac{2M_0^2}{\omega^2+4M_0^4},\\
 \pi_\eta(\omega)&=-\frac{\lambda}{6M_0^2}\frac{1}{\omega^2+4M_0^4}.
\end{align}

These equations are then solved as
\begin{align}
     I_\gamma(\omega) &=  \frac\lambda{6 M_0^2} \frac{2M_0^2(1+\tilde \lambda)}{\omega^2 + 4M_0^4(1+\tilde \lambda)^2},\\
     I_\eta(\omega) &= - \frac\lambda{6 M_0^2} \frac1{\omega^2 + 4M_0^4(1+\tilde \lambda)^2},
    \label{}
\end{align}
where $\tilde\lambda$ has been defined in Eq.~\eqref{eq:lambdatilde}. 
Inserting in the definition \eqref{appeq:defII} gives Eq.~\eqref{eq:IIsol}.

Turning now to the case of the $\ev{\chi\chi}$ correlator let us first prove Eq.~\eqref{eq:phi2phi2I}. In the symmetric phase, one has
\begin{align}
        \ev{\Phi_A\Phi_B\Phi_C\Phi_D}_{\rm nc} &= G_{AB}G_{CD} + \mathrm{perms.} \nonumber\\
        & - G_{AE}G_{BF}G_{CG}G_{DH}\Gamma^{(4)}_{EFGH},
    \label{eq:4phi}
\end{align}
where $\ev{\ldots}_{\rm nc}$ includes the disconnected contributions, perms. denotes the relevant permutations, and where the capital indices encompass the time variable, the Grassmann variable, and the field space index, and $G_{AB}=\ev{\Phi_A\Phi_B}$. We use the LO expression for the four point vertex function, which is schematically 
\begin{equation}
    \Gamma^{(4)}_{ABCD} = \frac{i\lambda}{3N} \qty[ \delta_{AB}\delta_{CD} \mathbb D_{AC} + \mathrm{perms.} ]
    \label{}
\end{equation}
\begin{equation}
    \mathbb D^{ab}_{12}(\omega)=\delta^{ab}\qty(\delta_{12} + i \mathbb I_{12}(\omega)).
\end{equation}
Inserting this in Eq.~\eqref{eq:4phi} and retaining only the LO terms, one obtains, after some simple algebra,
\begin{equation}
        C =-\frac{3}{\lambda N} \left(\Pi+i\Pi\star\Pi -\Pi\star\Pi\star \mathbb{I}\right)
 \end{equation}
 where $\Pi$ has been defined in Eq.~\eqref{eq:Pi} and where we have defined
\begin{equation}
 (A\star B)_{12}(\omega)=\int_3A_{13}(\omega) B_{32}(\omega).
\end{equation}
Using the defining equation \eqref{eq:integeq} for $\mathbb I$, 
\begin{equation}
 \mathbb I=\Pi+i\Pi\star\mathbb I,
\end{equation}
we finally get
 \begin{align}
      C=-\frac{3}{\lambda N} \left(\Pi+i\Pi\star \mathbb{I}\right)=-\frac{3}{\lambda N}  \mathbb{I}.
    \label{}
\end{align}

\end{document}